\documentclass[preprint,aps,superscriptaddress,nofootinbib,showkeys,11pt]{revtex4}
\pdfoutput=1
\usepackage{graphicx}
\usepackage{amsmath}
\usepackage{amsfonts}
\usepackage{amssymb}
\usepackage{xcolor, soul}
\usepackage{epstopdf}
\usepackage{float}
\usepackage{subfigure}
\usepackage{hyperref}
\hypersetup{
     colorlinks   = true,
     citecolor    = red,
     linkcolor    = blue,
     urlcolor     = blue,
}
\def\beq{\begin{equation}}
\def\eeq{\end{equation}}
\def\bea{\begin{eqnarray}}
\def\eea{\end{eqnarray}}
\def\be{\begin{equation}}
\def\ee{\end{equation}}
\def\pr{\partial}
\def\nno{\nonumber}
\def\bse{\begin{subequations}}
\def\ese{\end{subequations}}

\graphicspath{{./figs/}}
\begin{document}

\title{Probing the reheating phase through primordial magnetic field and CMB\\
}

\author{Md Riajul Haque}%
\email{riaju176121018@iitg.ac.in}
\author{Debaprasad Maity}
\email{debu@iitg.ac.in}
\author{Sourav Pal}%
\email{pal.sourav@iitg.ac.in}
\affiliation{%
	Department of Physics, Indian Institute of Technology Guwahati.\\
	Guwahati, Assam, India 
}%

\date{\today}

\begin{abstract}
Inflationary magnetogenesis has long been assumed to be the most promising mechanism for producing large-scale magnetic fields in our universe. However, generically, such models are plagued with either backreaction or strong coupling problems within the standard framework. This paper has shown that the reheating phase can play a crucial role in alleviating those problems along with CMB. Assuming the electrical conductivity to be negligible during the entire period of reheating, the classic Faraday electromagnetic induction changes the magnetic field's dynamics drastically. Our detailed analysis reveals that this physical phenomenon not only converts a large class of magnetogenesis model observationally viable without any theoretical problem but also can uniquely fix the perturbative average inflaton equation of state, $\omega_{\phi} = (p+2)/(p+2)$ during reheating given a specific value of the large scale magnetic field.
This observation hints the inflaton to assume the potential of form $V(\phi) \sim \phi^p$ near its minimum with $p \gtrsim 3.5$ if one considers the limit of the present day strength of the large scale magnetic field to be $\mathcal{P}_{B0}^{\frac{1}{2}} \gtrsim 10^{-18}$ G. 
Our analysis opens up a new avenue towards constraining the inflationary and magnetogenesis model together via reheating.        
\end{abstract}
\maketitle
\section{Introduction}
Reheating is one of the most important early phase of our universe. It essentially links the standard thermal universe with its pre-thermal phase namely the inflationary universe through a complicated non-linear process. Over the years major cosmological observations \cite{Akrami:2018odb, Aubourg:2014yra,Baumann:2008bn,Vazquez:2018qdg} have given us ample evidences in understanding the theoretical as well as observational aspects of both the thermal and the non-thermal inflationary universe cosmology to an unprecedented label. 
However, the intermediate reheating phase is still at its novice stage in terms of both theory and observation.   
From the cosmic microwave background
 (CMB) anisotropy \cite{Akrami:2018odb}, one can estimate the baryon content of the universe which agrees extremely well with the theoretical prediction of big-bang nucleosynthesis (BBN) \cite{Kawasaki:2000en, Steigman:2007xt,Fields:2014uja,Kawasaki:1999na}. Furthermore, with the successful standard big-bang model we have a very good understanding over a large time scale of the universe from the present  (redshift $z = 0$) to BBN stage $(z \sim 10^9)$ at an energy scale $\sim {\cal O}(1)$ MeV. The tiny fluctuation of CMB anisotropy can be successfully linked with the almost scale-invariant density fluctuation predicted by the inflation in the early Universe \cite{Riotto:2002yw,Baumann:2009ds}.
Therefore, precession CMB data has provided us significant insight into how our universe evolves during inflation.   
However, in this paper, our goal is to understand the intermediate phase which joins the end of inflation and the BBN. This phase is largely ill-understood due to lack of observational evidences. It is generically described by the coherently oscillating inflaton and its non-linear decay into the radiation field. In the Boltzmann description, the phase is parametrized by the reheating temperature $(T_{re})$ and the reheating equation of state ($\omega_{re}$). Still now both the parameters remain unconstrained except the reheating temperature which is approximately bounded within $10^{15} \mbox{GeV} > T_{re} > T_{BBN} \sim 10~ \mbox{MeV}$.
However, if one takes into account non-perturbative reheating in the beginning, the upper bound on reheating temperature could be within $10^{10} - 10^{13}$ GeV \cite{Haque:2020zco}.
Attempts to understand this phase in the literature can be broadly classified into two categories: i) Studying the background dynamics during reheating reveals useful information about the deep connection among the inflationary scalar spectral index $(n_s)$ and the reheating temperature $(T_{re})$, and reheating equation of state $(\omega_{re})$ \cite{Martin:2010kz,Dai:2014jja,Maity:2018dgy,Maity:2018qhi}. 
ii) Evolution of inflationary stochastic gravitational waves has been shown to encode valuable information when passing through this phase \cite{caprini,nakayama}.

In this paper we consider present-day Large Scale Magnetic Field (LSMF) combined with the CMB anisotropy to probe the reheating phase of the universe followed by the standard inflationary phase. For LSMF we consider simple model of primordial magnetogeneis  \cite{Kobayashi:2019uqs,Ferreira:2013sqa,ratra,Martin:2007ue}. While probing the reheating phase through those observables, we observe how the magnetogensis models itself will be constrained by the observables as well. Inflationary magnetogenesis models have been studied quite extensively in the literature \cite{subramanian,Martin:2007ue,Kobayashi:2014zza,Fujita:2016qab,Shtanov:2019civ,Guo:2015awg,turner,Bamba:2014vda}. Mechanisms known so far are to introduce the interaction Lagrangian which explicitly breaks the conformal invariance in the electromagnetic sector. However, this mechanism generically suffers from either strong coupling or backreaction problem \cite{Demozzi:2009fu,Sharma:2017eps} which will be elaborated as we go along.
In this regard reheating phase has recently been shown to play a very important \cite{Kobayashi:2019uqs} role.
As stated earlier the primary motivation of our present study will be to see in detail how the problems can be resolved partially by the reheating phase for various inflationary models, and simultaneously provide constraints on the reheating. Taking into account both the CMB anisotropic constraints on the inflationary power spectrum and the present value of the large scale magnetic field, our analysis reveals an important connection among the reheating parameters $(T_{re},w_{re})$, magnetogensis models and inflationary scalar spectral index $(n_s)$. 
 
The universe is observationally proved to be magnetized over a wide range of scales. Zeeman splitting, synchrotron emission, and Faraday rotation are some of the fundamental physical mechanisms by which the existence of a magnetic field can be probed. Various astrophysical and cosmological observations of those quantities tell us that our universe is magnetized over scales starting from our earth, the sun, stars, galaxies, galaxy
clusters, and also the intergalactic medium (IGM) in voids. In the galaxies and galaxy clusters of few to hundred-kilo parsecs (kpc) scale, the magnetic fields have been observed to be of order a few micro Gauss \cite{Grasso:2000wj, kronberg, Widrow:2002ud}. The $\gamma$-ray observations
of GRB 190114C, the observed long-term GeV-TeV light curve of the BLAZAR MRK 421 and BLAZAR MRK 501 suggest that even the intergalactic medium (IGM) in voids can host a weak $\sim 10^{-16}-10^{-20}$ Gauss magnetic field, with the coherence length as large as Mpc scales \cite{Wang:2020vyu,Takahashi:2011ac,Takahashi:2013lba, Asplund:2020frm, Kachelriess:2020bjl,Neronov:1900zz,Essey:2010nd,Finke:2015ona,Saga:2018ont}. Furthermore, future space based $\gamma-$ray observatories such as MAST with improved sensitivity will be able to probe the strength of the extra-galactic mangnetic field(EGMF) even below $B<10^{-18}$ G using the pair echo method \cite{Dzhatdoev:2020yvn}. Additionally, Jedamzik et al. \cite{Jedamzik:2018itu} derive an upper limit on the PMFs, which is 47 pG for scale-invariant PMFs, using MHD computation, which associates calculated CMB anisotropies with those observed by the WMAP
and Planck satellites. Therfore, all these efforts and obsevations are hinting towards the importance of understanding large scale magnetic field in the theoretical framework. In this context our main focus of the paper would be to understand the evolution of large scale magnetic field. We will not concentrate on the magnetogenesis scenario which has already been studied quite extentsively in the literature \cite{subramanian,Martin:2007ue,Kobayashi:2014zza,Fujita:2016qab,Guo:2015awg}. As the mechanism of inflationary magnetogensis is essentially the same, we will specifically consider the well known Ratra model \cite{ratra} for our study. Once the electro-magnetic field is generated during inflation, the subsequent evolution occurs through the time-evolving plasma. The magnetic field at the cosmological scale can evolve and survive even today while passing through the plasma state\cite{Boyarsky:2011uy,Matthews:2017apu} of our universe before the structure formation. The time-evolving plasma, therefore, proves to be an ideal environment to have a sustainable evolution and growth of the magnetic field. However, any physical processes responsible for the successful magnetogenesis inside the time-evolving plasma, it is the tiny seed initial magnetic field which plays a significant role. In the cosmological context, the most popular mechanism in this regard is the primordial inflationary magnetogenesis. Inflation provides us an outstanding mechanism for producing coherent magnetic fields for a wide range of scales. Mpc scale magnetic field can survive until today as a cosmological relic whose magnitude could be $\sim 10^{-9}-10^{-20}$ G. On the other hand at small scales, this tiny inflationary magnetic field can be the seed field which will be further enhanced to galactic scale $\mu$G  order magnetic field by the well known \textit{Galactic dynamo} mechanism\cite{kronberg,Dolginov:1994qu,Sol:1991wn,Brandenburg:1996fc,sur}. Similar kind of mechanism can also be obtained from the Electroweak Phase Transition (EWPT)  \cite{vega,banerjee,vachaspati,Zhang:2019vsb}.

We consider a standard scenario of inflationary magnetogenesis where electromagnetic field kinetic term is conformally coupled with a scalar field. Background inflation dynamics naturally produces a large-scale electromagnetic field which subsequently evolves through the reheating phase. Instead of going into the details of the magnetogenesis mechanism, we concentrate on dynamics during reheating considering various inflationary models. In our analysis we assume the negligible  Schwinger effect on the magnetogenesis. The paper is arranged as follows: in section \ref{sec2} we discuss the general analysis of the primordial magnetic fields from inflation and also the reheating dynamics used to constraint the parameters in the scenario of inflationary magnetogenesis. In section \ref{sec2} we also discuss different inflationary models and their dynamics, which are used to constrain the parameters. Subsequently in section \ref{sec3} we finally show how our analysis constrain the reheating as well as magnetogenesis model considering few observationally viable inflationary scenario.
 \section{Inflationary magnetogenesis: General discussion}\label{sec2}

During inflation the large scale magnetic field is generated out of quantum vacuum, and then subsequently evolves though various phases of our universe. Therefore, the evolving magnetic field must encode the valuable information about the reheating. Considering the present value of the large scale magnetic field we, therefore, can place constraints not only on the parameters of the reheating phase, but also on the magnetogensis model itself. Contrary to the convention, the important point of our present analysis is the assumption of  conductivity being negligible until the end of reheating. The reason being the production of the radiation plasma occurs nearly at the end of reheating. From large number of studies  \cite{Dai:2014jja,Martin:2010kz,Cook:2015vqa,Maity:2018dgy}, it is observed that almost entire period of reheating is primarily dominated by inflaton.
In the context of inflationary magnetogensis scenario, this particular assumption has recently been proposed to be important \cite{Kobayashi:2019uqs} during reheating.  
Conventionally after the end of inflation, the magnetic field on the super-horizon scales is assumed to be redshifted with the scale factor $a$ as $ B^2\propto {1}/{a^4} $ provided inflaton energy density transfers into plasma and the universe become good conductor instantly right after the end of inflation. Hence, the electric field ceases to exist. However, in the reference, \cite{Kobayashi:2019uqs} it has been shown that if the conductivity remains small redshifts of magnetic energy density becomes slower, $
B^2\propto {1}/{a^6H^2}$, due to electromagnetic Faraday induction. Here $H$ is the Hubble parameter. This helps one to obtain the required value of the present day large scale magnetic field considering large class inflationary model which we describe below,   
\subsection{Quantizing the Ratra model: Electromagnetic power spectrum}
The simplest magnetogenesis scenario which one can think of is the well known scalar-gauge field model with the following interaction Lagrangian $I(\phi)^2FF$, well known as Ratra model \cite{ratra}. In this interacting Lagrangian, the conformal symmetry is explicitly broken by the scalar field coupling function $I(\phi)$ in the gauge field sector.
During inflation, the model generally predicts strong primordial electric field than magnetic field, and that can backreact to invalidate the mechanism itself. First we discuss the model in detail and show how the reheating phase can come as our rescue of this backreaction problem. In the frame of a comoving observer having four-velocity $u^{\mu}$ ($u^i=0,u_{\mu}u^{\mu}=-1$), the magnetic and electric fields are defined as 
\bea \label{eq3}
E_{\mu}=u^{\nu} F_{\mu\nu}~,~ B_{\mu}=\frac{1}{2}\epsilon_{\mu\nu\rho\sigma}u^{\sigma}F^{\nu\rho}.
\eea
Where, $F_{\mu\nu}=\pr_{\mu}A_{\nu}-\pr_{\nu}A_{\mu}$ is the electromagnetic field tensor and $\epsilon_{\mu\nu\rho\sigma}$ is a totally antisymmetric tensor. 
The background is the well known  FLRW metric with the time dependent scale factor $a(\tau)$ expressed in conformal coordinate,
\beq
\label{eq4}
ds^2=a(\tau)^2\left(-d\tau^2+d\textbf{x}^2\right)~~.
\eeq
As already described before, the gauge field action is taken to be,
\beq
\label{eq5}
S=-\frac{1}{4}\int d^4x \sqrt{-g}I(\tau)^2F_{\mu \nu}F^{\mu \nu}~~,
\eeq
At this point let us point out that one can consider other inflationary magnetigeneis models with axion-electromagnetic field coupling \cite{Patel:2019isj,Domcke:2018eki,Adshead:2016iae}, higher curvature coupling \cite{Atmjeet:2013yta,Barrow:2012ty} and apply our methodology presented here to not only constraint the reheating phase but also make the models under consideration viable.

 In the inflationary magnetogenesis scenario, the essential idea is to quantize the electromagnetic field in the classical inflationary background. Here $I(\tau)^2 \equiv I(\phi(\tau))$, therefore, is the time dependent coupling arising form some classical background scalar field. To maintain generality we do not specify any background dynamics of the scalar field. Through this coupling the electromagnetic field experiences the spatially flat expanding FLRW background. In order to quantize the field components $A_{\mu}$ are expressed in terms of irreducible scalar and vector components as follows,  
\beq
\label{eq6}
A_\mu=\left(A_0,\partial_iS+v_i\right)~~ \mbox{with}~~ \partial_iV_i=0~~.
\eeq 
In the standard canonical quantization procedure, one writes $V_i$ in terms of the annihilation ($a_k$) and the creation operator ($a^\dagger_k$) as
\beq
\label{eq7}
V_i(\tau,x)=\sum_{p=1,2}\int\frac{d^3k}{(2\pi)^3}\epsilon^{(p)}_i(\textbf{k})\left\lbrace e^{i\textbf{k.x}}a_k^{(p)}u_k^{(p)}(\tau)+e^{-i\textbf{k.x}}a_k^{\dagger(p)}u_k^{*(p)}(\tau)\right\rbrace~~,
\eeq
here the $\epsilon_i^{(p)}(k)$ is the polarization vector corresponding to the two polarization direction $p=1,2$, which satisfy the following relations, $
\epsilon_i^{(p)}(\textbf{k}) k_i=0 ~,~  \epsilon_i^{(p)}(\textbf{k})\epsilon_i^{(q)}(\textbf{k})=\delta_{pq}~~.
$
The creation and annihilation operators in Eq.\eqref{eq7} namely $a_k^{(p)}$ and $a_k^{\dagger(p)}$ are time independent. They satisfy the commutation relation 
\beq
\label{eq9}
[a_k^{(p)},a_h^{(q)}]= [a_k^{\dagger(p)},a_h^{\dagger(q)}] =0 ~,~ [a_k^{(p)},a_h^{\dagger(q)}]=(2\pi )^3\delta^{pq}\delta^{(3)}(\textbf{k-h})~~.
\eeq
All the dynamics of the field will be encoded into the mode function which satisfy the following equation of motion, 
\beq
\label{eq10}
u^{(p)\prime\prime}_k+2\frac{I^\prime}{I}+k^2u^{(p)\prime}_k=0 .
\eeq
Where the prime denotes derivative with respect to the proper time $\tau$.
Conventionally the electromagnetic power spectrum is expressed in terms of those mode functions as follows,
\beq
\label{eq11}
\textit{P}_E(k)=\frac{k^3}{2\pi^2a^4}\sum_{p=1,2}\vert u^{(p)\prime}_k\vert^2~;~\textit{P}_B(k)=\frac{k^5}{2\pi^2a^4}\sum_{p=1,2}\vert u^{(p)}_k\vert^2.
\eeq
However, it would look physically elegant, if we express the power spectrum in terms Bogliubov coefficient which has direct physical interpretation in terms of quantum particle production. In order to do that let us first write down the Hamiltonian in terms of time independent creation and annihilation operator as follows,  
\bea \label{eq12}
H=\frac{I^2(\tau)}{2}\sum_{p=1,2}\left(k\vert u^{(p)}_k\vert^2+\frac{1}{k}\vert u^{\prime(p)}_k\vert^2\right)\left(a_k^{(p)\dagger} a^{(p)}_k + a^{(p)}_{-k} a_{-k}^{(p)\dagger}\right) +\left(k (u^{(p)}_k)^2+\frac{1}{k}(u^{\prime(p)}_k)^2\right)a^{(p)}_k a^{(p)}_{-k} \nonumber\\
+\left(k (u^{(p)*}_k)^2+\frac{1}{k}(u^{*\prime(p)}_k)^2\right) a_{-k}^{(p)\dagger} a^{(p)\dagger}_k
+\frac{1}{2}\left(a_k^{(p)\dagger} a^{(p)}_k - a^{(p)}_{-k}a^{(p)\dagger}_{-k}\right) .
\eea
This is clearly not diagonal. Therefore, in order to diagonalize, we employ the Bogoliubov transformation.
In this transformation, new set of time dependent creation and annihilation operators $(b^{(p)}_k(\tau),b_k^{(p)\dagger}(\tau))$ are defined in term of old ones. And the new basis of the Hilbert space so constructed  diagonalizes the above Hamiltonian, 
\bea \label{eq13}
b^{(p)}_k(\tau)=\alpha_k^{(p)}(\tau)a^{(p)}_k+\beta_k^{(p)*}a_{-k}^{(p)\dagger}~,~ b_k^{(p)\dagger}(\tau)=\alpha^{(p)*}_k(\tau)a^{(p)\dagger}_k+\beta_k^{(p)}(\tau)a^{(p)}_{-k} .
\eea 
Where $\alpha^{(p)}_k$ and $\beta^{(p)}_k$ are the Bogoliubov coefficients defined as
\bea \label{eq14}
\alpha^{(p)}_k(\tau)=I\left(\sqrt{\frac{k}{2}}u^{(p)}_k+\frac{i}{\sqrt{2k}}u_k^{\prime(p)}\right)\\
\beta^{(p)}_k(\tau)=I\left(\sqrt{\frac{k}{2}}u^{(p)}_k-\frac{i}{\sqrt{2k}}u_k^{\prime(p)}\right) .
\eea 
The Bogoliubov coefficients follow the following normalization condition 
\beq
\label{eq15}
\vert\alpha_k^{(p)}\vert^2-\vert\beta_k^{(p)}\vert^2=1
\eeq
With all these ingredients one represents the power spectrum in terms of Bogoliubov coefficients, $\alpha^{(p)}_k$ and $\beta_k^{(p)}$ as follows,
\beq \label{eq16}
\mathcal{P}_E(k)=\frac{k^4}{4\pi^2a^4I^2}\sum_{p=1,2} \vert\alpha^{(p)}_k-\beta^{(p)}_k\vert^2~;~\mathcal{P}_B(k)=\frac{k^4}{4\pi^2a^4I^2}\sum_{p=1,2}\vert\alpha^{(p)}_k+\beta^{(p)}_k\vert^2
\eeq
Considering the super-horizon limit, the above expressions for the electromagnetic power spectrum can be further simplified by extracting the amplitude and phase of those coefficients. By using the normalization condition Eq \eqref{eq15} one can write,
\beq
\label{eq17}
\vert\alpha_k^{(p)}\pm \beta^{(p)}_k\vert^2=1+2\vert\beta_k^{(p)}\vert^2\pm\vert\beta_k^{(p)}\vert\sqrt{1+\vert\beta_k^{(p)}\vert^2}\cos\lbrace arg(\alpha^{(p)}_k\beta_k^{*(p)})\rbrace
\eeq
Where, the phase factor is expressed as $arg(\alpha^{(p)}_k\beta_k^{*(p)})\equiv\pi+\theta^{(p)}_k$. In the following discussion, we consider a specific model of the electromagnetic coupling function.
\subsubsection*{\textbf{Magnetogenesis: Modelling the coupling function}}
In order to study further we consider the following widely considered power law form of the coupling function \cite{ratra} 
\bea
\label{eq21}
I(\tau)=\left\{
\begin{array}{ll}
	\left(\frac{a_{end}}{a}\right)^n \qquad a\leq a_{end}\\	
	1 \hspace{53pt}a\geq a_{end} ,
\end{array} \right.
\eea
where, $a_{end}$ is the scale factor at the end of inflation. At this stage let us point out that, in all the previous analysis in the literature, $n$ has been considered to be integer. However, for our present discussion we keep the value of $n$ arbitrary. As one of the important focuses of the analysis also is to constrain the magnetogenesis model itself namely the value of $n$ in accord with the present day large scale magnetic field and CMB anisotropy. Further more after the end of inlation the value of the function is so chosen that the usual conformal electrodynamics is restored. 

Our main interest is to understand the large scale magnetic field which is assumed to be produced during the initial stage of inflation. Therefore, through out our analysis, we  assume the Hubble parameter to be constant. This also helps us to give clear picture in terms of analytic solutions. The perfect de-Sitter background with Hubble parameter $H_{inf}$,
one obtains the solution for the mode function as,  
\beq
\label{eq22}
u_k=\frac{1}{2I}\left(\frac{\pi}{a H_{inf}}\right)^{\frac 12}H^{(1)}_{-n+\frac{1}{2}}\left(\frac{k}{aH_{inf}}\right),
\eeq
which leads to Bunch-Davis vacuum state at the sub-Horizon scale.   $H^{(1)}_{-n+\frac{1}{2}}\left(\frac{k}{aH_{inf}}\right)$ is the Hankel function of first kind. The Hankel functions are defined in terms of the Bessel function of first and second (Neumann) kind.
\beq
\label{eq23}
H_{\nu}\left(\frac{k}{aH_{inf}}\right)=J_{\nu}\left(\frac{k}{aH_{inf}}\right)+iY_{\nu}\left(\frac{k}{aH_{inf}}\right)
\eeq
Here $\nu=-n\pm\frac{1}{2}$.
Defining a new variable $z = {k}/{a H_{inf}}$, the time dependent Bogoliubov coefficient are expressed as, 
\beq
\label{eq25}
\alpha_{k}=\left(\frac{\pi z}{8}\right)^{\frac{1}{2}}\left\lbrace H^{(1)}_{-n+\frac{1}{2}}(z)-i H^{(1)}_{-n-\frac{1}{2}}(z)\right\rbrace~;~
\beta_{k}=\left(\frac{\pi z}{8}\right)^{\frac{1}{2}}\left\lbrace H^{(1)}_{-n+\frac{1}{2}}(z)+i H^{(1)}_{-n-\frac{1}{2}}(z)\right\rbrace
\eeq
Now as we have already described before, our focus is to understand the reheating constraints on inflationary as well as magnetogenesis models considering the CMB anisotropy and LSMF field, which are naturally classified as large scale observables. Our discussion will be mostly in the super-horizon limit. Therefor, in this limit following condition will be satisfied
\beq
\label{eq19}
\frac{1}{\vert\beta^{(p)}_k\vert^2}\ll\theta^{(p)}_k\ll1 .
\eeq
This essentially suggests that any mode starting from the Bunch-Davis vacuum at the sub-horizon scale transforms into highly squeezed state, parametrized by $\vert\beta^{(p)}_k\vert^2 \gg 1$, after its horizon exit during inflation. Using this condition one gets the following simplified form of the electromagnetic power spectrum, 
\beq \label{eq20}
\mathcal{P}_E(k)\simeq \frac{k^4}{4\pi^2a^4I^2}\sum_{p=1,2}4\vert\beta^{(p)}_k\vert^2~;~ \mathcal{P}_B(k)\simeq \frac{k^4}{4\pi^2a^4I^2}\sum_{p=1,2}\vert\beta^{(p)}_k\vert^2\left(\theta^{(p)}_k\right)^2
\eeq
For any arbitrary value of $n$, the final expression for the spectrum are as follows,
\bea
\label{eq26}
\mathcal{P}_E(k)\simeq \frac{4k^4}{2\pi^2a^4I^2}\left(\frac{\pi z}{8}\right)\left\lbrace H^{*(1)}_{-n+\frac{1}{2}}(z)H^{(1)}_{-n+\frac{1}{2}}(z)+iH^{*(1)}_{-n+\frac{1}{2}}(z)H^{(1)}_{-n-\frac{1}{2}}(z)\right\rbrace\nonumber\\
-\frac{4k^4}{2\pi^2a^4I^2}\left(\frac{\pi z}{8}\right)\left\lbrace i H^{*(1)}_{-n-\frac{1}{2}}(z)H^{(1)}_{-n+\frac{1}{2}}(z)-H^{(1)}_{-n-\frac{1}{2}}(z)H^{*(1)}_{-n-\frac{1}{2}}(z)\right\rbrace
\eea
\bea\label{eq27}
\mathcal{P}_B(k)\simeq \frac{k^4}{2\pi^2a^4I^2}\left(\frac{\pi z}{8}\right)\left\lbrace H^{*(1)}_{-n+\frac{1}{2}}(z)H^{(1)}_{-n+\frac{1}{2}}(z)+iH^{*(1)}_{-n+\frac{1}{2}}(z)H^{(1)}_{-n-\frac{1}{2}}(z)\right\rbrace\nonumber\\
-\frac{k^4}{2\pi^2a^4I^2}\left(\frac{\pi z}{8}\right)\left\lbrace i H^{*(1)}_{-n-\frac{1}{2}}(z)H^{(1)}_{-n+\frac{1}{2}}(z)-H^{(1)}_{-n-\frac{1}{2}}(z)H^{*(1)}_{-n-\frac{1}{2}}(z)\right\rbrace\nonumber\\
\left\lbrace arg\left[\frac{\pi z}{8}\left(H^{(1)}_{-n+\frac{1}{2}}(z)-iH^{(1)}_{-n-\frac{1}{2}}(z)\right)\left(H^{*(1)}_{-n+\frac{1}{2}}(z)-i H^{*(1)}_{-n-\frac{1}{2}}(z)\right)\right]-\pi\right\rbrace^2
\eea
In our final analysis, we will be considering this expression. To this end it is important to remind the reader, the widely studied case for $n$ being positive integer for which the electromagnetic power spectra assumes the following simple forms in the superhorizon limit ($k\ll aH_{inf}$) as \cite{Kobayashi:2019uqs}.
\beq
\label{eq28}
\mathcal{P}_E(k)\simeq\frac{8\Gamma(n+\frac{1}{2})^2 H^4_{inf}}{I^2\pi^3}\left(\frac{k}{2aH_{inf}}\right)^{-2(n-2)}~~;~~
\mathcal{P}_B(k)\simeq\frac{8\Gamma(n-\frac{1}{2})^2 H_{inf}^4}{I^2\pi^3}\left(\frac{k}{2aH_{inf}}\right)^{-2(n-3)} .
\eeq
With this expression, subsequent magnetic field evolution has been widely studied considering the magnetic energy density decreasing as $|B|^2 \sim 1/a^4$. The required value of large scale magnetic field of order $10^{-16}$ G can be obtained only if inflation scale $H_{inf}$ assumes low value which has been proved to be difficult in conventional inflationary model in the effective theory framework. We also have observed this in our numerical analysis. Simultaneously we also observed how this problem can be alleviated by the electromagnetic induction during reheating.
Furthermore, we will assume arbitrary value of $n$. 
\subsubsection*{\textbf{After inflation dynamics: reheating}}
In order to associate the observed current magnetic field with the magnetic field produced during inflation, it is essential to study the subsequent evolution. Most of the studies so far considered the fact that when $I^2$ becomes constant at the end of the inflation, the co-moving photon density $\vert\beta_k^{(p)}\vert^2$ is conserved. Consequently the magnetic power redshifts as $\mathcal{P}_B(k)\propto a^{-4}$ until today. Before we embark on our original analysis, for the sake of completeness let us briefly discuss this widely studied case. Thus at the end of the inflation, the phase parameter ($\theta_k^{end}$) and the photon density ($\vert\beta_k^{end}\vert^2$) are identified as
\bea
\theta_k^{end}= \left\lbrace Arg\left[\alpha_k\left(z_{end}\right)\beta_k^{*}(z_{end})\right]-\pi\right\rbrace~~,
\eea
\bea\label{photonden}
\vert\beta_k^{end}\vert^2=\left(\frac{\pi z_{end}}{8}\right)\left\lbrace H^{*(1)}_{-n+\frac{1}{2}}(z_{end})H^{(1)}_{-n+\frac{1}{2}}(z_{end})+iH^{*(1)}_{-n+\frac{1}{2}}(z_{end})H^{(1)}_{-n-\frac{1}{2}}(z_{end})\right\rbrace\nonumber\\
-\left(\frac{\pi z_{end}}{8}\right)\left\lbrace i H^{*(1)}_{-n-\frac{1}{2}}(z_{end})H^{(1)}_{-n+\frac{1}{2}}(z_{end})-H^{(1)}_{-n-\frac{1}{2}}(z_{end})H^{*(1)}_{-n-\frac{1}{2}}(z_{end})\right\rbrace~~.
\eea
Here we define $z_{end}\equiv {k}/{a_{end}H_{inf}}$. For any arbitrary values of $n$, the magnetic power spectrum is
\bea\label{magetic1}
\mathcal{P}_B^{end}(k)\simeq \frac{k^4}{2\pi^2 a_{end}^4}\left(\theta_k^{end}\right)^2\vert\beta_k^{end}\vert^2~~.
\eea
For positive integer $n$, the expression above (\ref{magetic1}) boils down to
\bea\label{magnetic2}
\mathcal{P}_B^{end}(k)\simeq \frac{8\Gamma(n-\frac{1}{2})^2}{\pi^3}H_{Inf}^4\left(\frac{k}{2a_{end}H_{Inf}}\right)^{-2(n-3)}~~,
\eea
at the super-horizon scale $k\ll a H_{inf}$.
Considering nontrivial dynamics, the magnetic power spectrum at the present universe can be correlated with the magnetic power at the end of the inflation through the following standard equation
\bea \label{magnetic3}
\mathcal{P}_{B0}(k)\simeq \mathcal{P}_B^{end}(k)\left(\frac{a_{end}}{a_0}\right)^4~~.
\eea
Where ${a_{end}}/{a_0}$ can be expressed in terms of inflationary e-folding number ($N_k$) as
\bea\label{magnetic4}
\frac{a_{end}}{a_0}=\frac{a_{end}}{a_k}\frac{k'}{a_0H_{Inf}}=\left(\frac{k'}{a_0}\right)\frac{e^{N_k}}{H_{Inf}}~~.
\eea
Here ${k'}/{a_0}=0.05 M_{pc}^{-1}$ is taken as pivot scale set by Planck observation. By combining equations (\ref{magnetic2}), (\ref{magnetic3}), and (\ref{magnetic4}), the expression for the magnetic power considering integer values of $n$, follow the  equation
\bea
\mathcal{P}_{B0}\simeq \frac{\Gamma(n-\frac{1}{2})^2}{2^{3-2n}\pi^3}\left(\frac{k}{a_0}\right)^{6-2n}\left(\frac{k'}{a_0}\right)^{2n-2}e^{2(n-1)N_k}~~.
\eea
\begin{figure}[t!]
 	\begin{center}
 		\includegraphics[width=009.00cm,height=6.3cm]{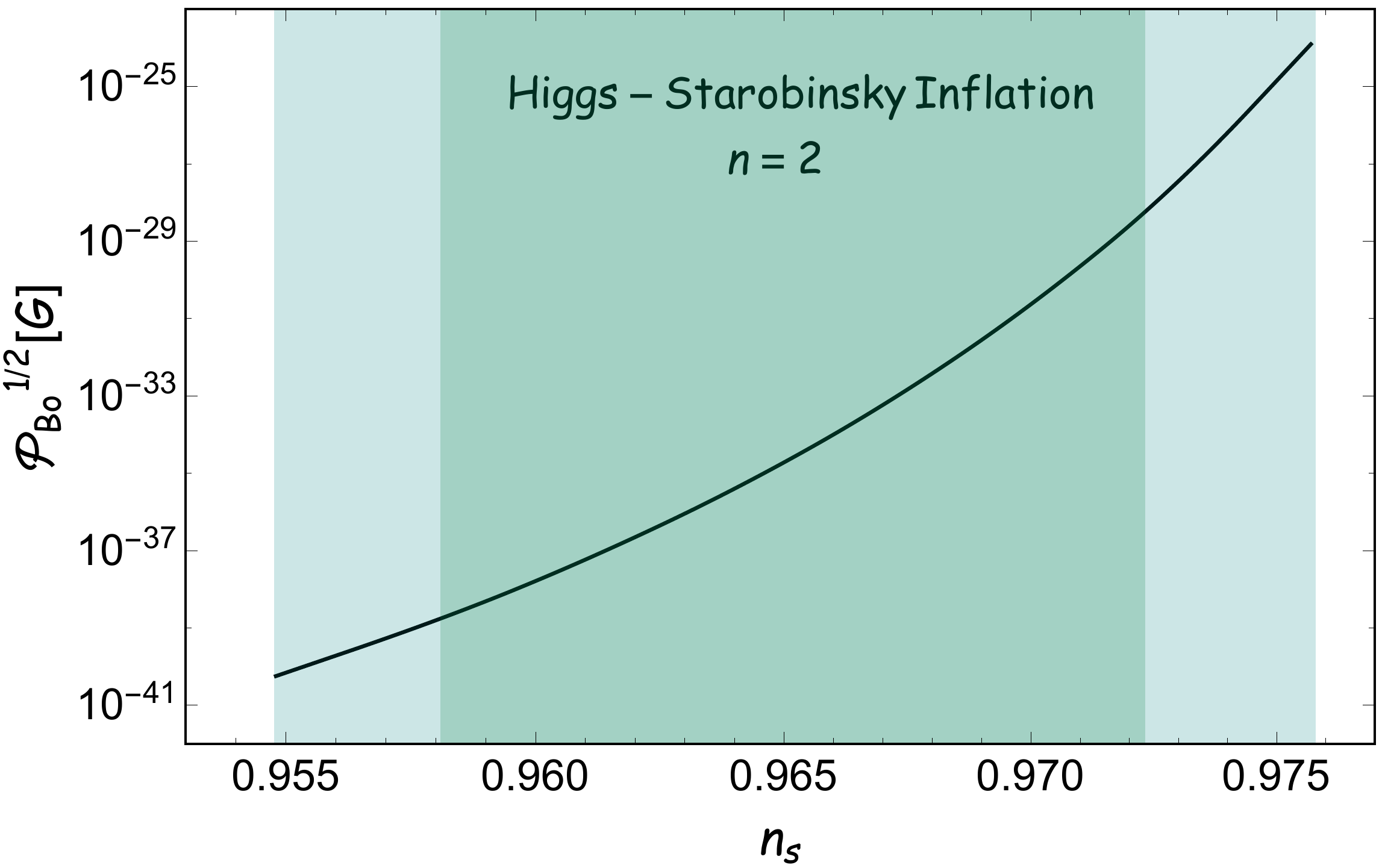}
 		
 		\caption{We plot the variation of the present magnetic field's strength as a function of the spectral index within 2$\sigma$ range of $n_s$ from Planck \cite{Akrami:2018odb} for Higgs-Starobinsky inflationary model. It is clear that the required magnetic field strength $(10^{-9}\sim 10^{-22})$ G is difficult to achieve within the conventional framework, unless one introduces slow decreasing rate of magnetic energy density in some early state of universe evolution.}
 		\label{strinf}
 	\end{center}
 \end{figure}
It has already been observed and also shown in Fig.\ref{strinf} the well known fact that magnetic strength of order $10^{-9}-10^{-22}$ Gauss on 1 Mpc scales \cite{Martin:2007ue,Neronov:2013zka,Takahashi:2013lba,Vachaspati:2016xji} can not be obtained by the above magnetic power spectrum for  high-scale inflation model such as well studied Higgs-Starobinsky inflation. Therefore, number of models have been constructed just to avoid this problem without much success with regard to the theoretical issues which we discuss in the next section. Form here itself we will advocate the need for re-looking into the reheating effect more seriously than model building.  In the recent paper \cite{Kobayashi:2019uqs}, authors have showed that considering negligible electrical conductivity during reheating the Faraday's law of electromagnetic induction plays an interesting role in modifying the magnetic field evolution. Therefore, we want to see how this induction effect  can safely generate presently observed magnetic field for different high scale inflationary models. Interestingly incorporating the CMB anisotropy into the analysis will further reveal an intricate interconnection among various apparently disconnected cosmological parameters such as $(n_s, P_{B0}, T_{re}, w_{re})$. Our analysis, therefor, opens up an interesting new possibility of probing the reheating dynamics, which is otherwise difficult, and simultaneously constraining the inflationary as well as magnetogenesis model parameters through the evolution of the primordial magnetic field.\\
After the end of inflation the gauge kinetic function $I(\phi)$ is assumed to be unity, rendering the fact that the post inflationary evolution of the electromagnetic field is essentially standard Maxwellian. Therefore, the conformal invariance is restored, and consequently the gauge field production from the quantum vacuum ceases to exist. The electromagnetic field produced during inflation will cross the horizon and turned into the classical one which will subsequently evolve during this phase.
In the Fourier space, the mode function solution of the free Maxwell equation is,
\bea
\label{eq29}
u_k^{(p)}=\frac{1}{\sqrt{2k}}\lbrace\alpha_k^{(p)}(z_{end})e^{-ik(\tau-\tau_{end})}+\beta_k^{(p)}(z_{end})e^{-ik(\tau-\tau_{end})}\rbrace~,
\eea
here $\tau_{end}$ represents the end of inflationary era. For arbitrary value of $n$, the phase factor is calculated to be 
\bea
\label{eq30}
\theta_k^{(p)}= \left\lbrace Arg\left[\alpha_k^{(p)}\left(z_{end}\right)\beta_k^{(p)*}(z_{end})\right]-\pi\right\rbrace-2k\left(\tau-\tau_{end}\right)~.
\eea
Where the elapsed conformal time is obtained as
\bea
\tau-\tau_{end}=\int_{a_{end}}^a \frac{da}{a^2H}~~.
\eea
Since there is no further production of gauge field, the photon number density $\vert\beta_k\vert^2$ becomes independent of time and follows the same equation (\ref{photonden}) as before. However, as emphasized already, the Farady induction will come into play during this phase. In order to see this let us first express the  magnetic power spectra at the end of reheating parametrized by the scale factor ($a_{re}$) as
\bea
\label{eq32}
\mathcal{P}_{B_{re}}(k)\simeq\frac{k^4}{2\pi^2 a^4_{re}}(\theta_k^{re})^2\vert\beta_k^{end}\vert^2~~,
\eea
where $\theta_k^{re}$ is the phase parameter at the end of the reheating era, which is defined as,
\bea\label{eq33}
\theta_k^{re}= \left\lbrace Arg\left[\alpha_k\left(z_{end}\right)\beta_k^{*}(z_{end})\right]-\pi\right\rbrace-2 k\int_{a_{end}}^{a_{re}}\frac{da}{a^2H}~.
\eea
Notable term of the above expression is the second one which leads to the non-conventional dynamics of the magnetic field. Assuming the  constant equation of state during reheating dynamics the special term boils down to the following simple from,
\bea
2k(\tau-\tau_{end})\to\frac{4k}{3\omega+1}\left(\frac{1}{aH}-\frac{1}{a_{end}H_{inf}}\right) .
\eea
The phase term contributes to the dynamics of the magnetic power spectrum in two different ways. The conventional one will be associated with the redshift factor $\propto a^{-4}$ emerging from the first term in the right-hand side of the Eq.(\ref{eq30}). Important one is associated with the redshift factor $\propto a^{-6}H^{-2}$ emerged out  from ${k}/{aH}$ term.  As the expansion of the universe is decelerating after inflation, leading contribution to the evolution of the magnetic power would be controlled by the latter one Eq.(\ref{eq30}).
However, the electric field energy dilutes following the conventional form
\bea\label{electricre}
\mathcal{P}_{E}(k)\simeq\frac{2k^4}{\pi^2 a^4}\vert\beta_k^{end}\vert^2~~.
\eea
For large scale ($k\ll aH$), the above expression boils down to,
\bea
\mathcal{P}_E(k)\simeq \frac{8\Gamma(n+\frac{1}{2})^2}{\pi^3}H_{inf}^4\left(\frac{k}{2a_{end}H_{Inf}}\right)^{-2(n-2)}\left(\frac{a_{end}}{a}\right)^4~~.
 \eea
for integer values of $n$.
After the end of reheating, inflaton energy is converted into highly  conducting plasma containing all the standard model particels. Due to large electrical conductivity primordial electric field decays to zero and comoving magnetic energy density freezes to a constant value until today. Therefore, final general expression of our interest is the present day magnetic field strength given as
\bea
\label{eq34}
\mathcal{P}_{B_0}(k)=\mathcal{P}_{B_{re}}(k)\left(\frac{a_{re}}{a_0}\right)^4
\simeq\frac{k^4}{2\pi^2}(\theta_k^{re})^2\vert\beta_k^{end}\vert^2\frac{1}{a_0^4}~~.
\eea
If we take integer value of $n$, at super horizon scale, the above expression will transforms into   \cite{Kobayashi:2019uqs}
\bea \label{eq35}
\mathcal{P}_{B0}(k)\simeq \frac{8\Gamma(n-\frac{1}{2})^2}{\pi^3}H_{inf}^4\left(\frac{k}{2a_{end}H_{inf}}\right)^{-2(n-3)}\left(\frac{a_{end}}{a_0}\right)^4\nonumber\\
\left\lbrace 1+\left(2n-1\right)\int^{a_{re}}_{a_{end}}\frac{da}{a}\frac{a_{end}H_{inf}}{aH}\right\rbrace^2~,
\eea
where $a_{0}$ is the scale factor at the present time. In the next section we will briefly discuss about standard theoretical issues related to the magnetogenesis model.
\subsection*{\textbf{Discussion on strong coupling and backreaction problem}}\label{backreactionstrongcoupling}
In order to generate large scale magnetic field of required strength, conventional magnetogenesis models either encounter strong coupling or backreaction problem. In order to ameliorate these issue several attempts \cite{Ferreira:2013sqa,Demozzi:2009fu,Fujita:2012rb,Ferreira:2014hma,Green:2015fss,Ferreira:2015omg,Gasperini:1995dh,Bamba:2003av,Kanno:2009ei,Barnaby:2012tk,Suyama:2012wh,Jain:2012ga,Jain:2012vm,Nurmi:2013gpa,Fujita:2014sna,Ganc:2014wia} have been made either by advocating different forms of the gauge coupling function or taking non-trivial dynamics of the coupling fields during reheating. In the context of simplest magnetogenesis model proposed in \cite{ratra}, the gauge kinetic function $I(\tau)$ can essentially be interpreted as time dependent effective electromagnetic coupling. By considering field theoretic argument and experimental observations \cite{Demozzi:2009fu,Sharma:2017eps} it is generically argued that the model either suffers from strong coupling problem or backreaction problem. \\
In our previous discussion, we have chosen the effective electromagnetic coupling function as a monomial function of the scale factor,
\bea
I=\left(\frac{a_{end}}{a}\right)^{n}=e^{N_kn}~~,
\eea
where $N_k = \ln (a/a_{end})$ associated with a particular scale $k$ is identified as the e-folding number during inflation. It is clear from the above expression that once the value of $n$ is chosen to be negative, the gauge kinetic function increases during inflation. Now the behavior of $I(\tau)$ is so chosen that it boils down to unity after inflation. Therefore, during inflation, its magnitude must be less than unity. This is where the origin of the strong coupling problem in the electromagnetic sector lies. Under the field redefinition $A_{\mu} \rightarrow \sqrt{I} A_{\mu}$, the  effective electromagnetic coupling $\alpha_{eff}$, defined through the fermion-gauge field interaction $
{\cal L}_{int} = e A_{\mu} J^{\mu}
$ modified as 
\bea
\alpha_{eff} = \frac{e^2}{4 \pi I} = \frac{\alpha}{I}.
\eea
$e$ is the charge of the fermionic field contributing to the electric current. $\alpha$ is the standard electromagnetic coupling. Nonetheless, as one goes early in the inflationary phase, the effective electromagnetic coupling necessarily becomes very large, which turns the theory non-perturbative. Moreover, the perturbative computation of magnetogenesis, discussed in our previous section, will no longer be valid. Keeping this problem aside, if we still do our magnetogenesis computation as before, due to large effective electromagnetic coupling $\alpha_{eff} $, the production of electromagnetic energy density may be suppressed compared to the background energy density ($3M_p^2H_{Inf}^2$). Therefore, backreaction problem will not come into play.

 On the other hand, if one considers positive values of $n\ge 0$, the whole reasoning expressed just now will be reversed or, more precisely, throughout the inflationary period, the gauge kinetic function $I(\tau)$ will be way larger than unity. Based on our previous discussions, the perturbative magnetogenesis analysis appears to be still valid. However, considerable reduction of the effective electromagnetic coupling $\alpha_{eff} \propto 1/I(\tau)$, specifically during the early inflationary stage of our interest, significantly enhances the production of electromagnetic energy. In such a scenario the quantum production of electromagnetic energy may take over the background energy density. This is precisely the backreaction problem, which jeopardies the fixed background magnetogenesis analysis instead. 
To avoid the backreaction problem, the energy density of the gauge field must be smaller than the total background energy density  $\rho_{tot}=3M_p^2H_{Inf}^2$. The parameter which quantifies the amount of background energy density during inflation against the gauge field energy density $\rho_A$ is the ratio,
\bea\label{backr}
\frac{\rho_A}{\rho_{tot}}\leq \zeta~~.
\eea
Where $\zeta$ is identified with another physical quantity associated with the amplitude of the curvature perturbation measured from CMB anisotropy. From the CMB analysis the amplitude of the curvature perturbation is measured to be, $\zeta=4.58\times10^{-5}$ from Planck \cite{Akrami:2018odb}. The total gauge field energy density at a given scale factor $a_T$ during inflation can be calculated by using equations (\ref{eq16}) and (\ref{eq25}) as
\begin{eqnarray}\label{backr2}
\rho_A(a_T)&=&\rho_E(a_T)+\rho_B(a_T)=\frac{I(a_T)^2}{2} \int_{k_{IR}}^{k_T}\frac{dk}{ k }\left\{\mathcal{P}_E (k,a_T)+\mathcal{P}_B (k,a_T)\right\}\nno\\
&=&\frac{I(a_T)^2}{2} \int_{k_{IR}}^{k_T}\frac{dk}{4\pi k H_{end}}\left(\frac{k}{a_T}\right)^5 \left\{\left| H_{-n-\frac{1}{2}}^{(1)}\left(\frac{k}{a_T H_{end}}\right) \right| ^2+\left| H_{-n+\frac{1}{2}}^{(1)}\left(\frac{k}{a_T H_{end}}\right) \right| ^2\right\} .
\end{eqnarray}

In the above expression, the $k$-integration ranges from $k_{IR}$ to $k_{T}$. Associated with our observable universe, the IR cut-off $k_{IR}$ corresponds to the highest mode that exits the horizon at the beginning of the inflation or approximately the CMB scale and $k_T\sim a_T H_{end}$ denotes the mode that crosses the horizon at any arbitrary scale $a=a_T$.  In order to avoid backreaction, the energy density across the scales emerging during inflation ($a_k\leq a_T\leq a_{end}$), should satisfies the condition (\ref{backr}). So we can set a limit on the coupling parameter as $n_{max}$ above which the constraints equation (\ref{backr}) will be violated. Thus considering the coupling parameter within the aforementioned range, we are able to take care of both strong coupling and backreaction problems.\\
For example the scale-invariant electric power spectrum which corresponds to $n=2$, the total gauge field energy density at the end of the inflation is
\bea \label{scale2}
\rho_A=\rho_E+\rho_B \sim \frac{9}{4} \frac{H_{end}^4}{\pi^2} N_k+\frac{H_{end}^4}{8 \pi^2}\left(1-e^{-2N_k}\right)~~.
\eea
The ratio between the gauge field and the total background energy density is always less than the amplitude of the curvature perturbation (${\rho_A}/{\rho_{tot}}\sim 10^{-10}$) for all allowed values of the spectral index. For the scale-invariant electric power spectra model, therefore, the back reaction as well as strong coupling problem can be avoided. However, during entire inflation period such models generically produce magnetic field of insufficient strength. And we will see how the reheating phase helps to enhance the magnitude of the magnetic field to a desirable strength.  

Another interesting case is when the generated magnetic field is scale invariant which corresponds to $n=3$.
However, immediate problem arises in the electric field power spectrum $\mathcal{P}_E(k)\propto\left( \frac{k}{a H_{end}}\right)^{-2}\to \infty$ (see equation \ref{eq28}), which increases rapidly in the large scale limit $\frac{k}{a H_{end}}\to 0$. Hence electrical energy density exceeds the background inflaton energy density much before the end of the inflation. However,
it can be cured by changing the coupling function as follows,
 \bea
\label{eq21}
I(\tau)=\left\{
\begin{array}{ll}
	\left(\frac{a_{br}}{a}\right)^n \qquad a\leq a_{br}\\	
	1 \hspace{53pt}a\geq a_{br} ,
\end{array} \right.
\eea
where $a_{br}$ is the scale factor defined at a particular point during inflation when
\bea 
\frac{\rho_E(a_{br})+\rho_B(a_{br})}{3M_p^2H_{Inf}^2}=\zeta~~.
\eea
Therefore, the primary assumption is that the standard maxwell theory is recovered at a point $a_{br}$ not after the end inflation.
In such a scenario, one can naturally solve the backreaction problem for the $n=3$ magnetogenesis model. However, from detailed analysis we found it to be very difficult to obtain magnetic field at present time within the observable limit. The magnetic field strength turns out to be within $(10^{-45} - 10^{-50})$ G, which is very small compared with the observational limit. We will take up this issue in our future work. Therefore, our subsequent discussion will be mostly concentrated on the model with scale invariant electric power spectrum during inflation. 
\subsection{Reheating dynamics: Connecting Reheating and Primordial magnetic field via CMB}\label{sec2B}
By now it becomes clear that the primary importance of the reheating phase is to enhance the strength of the large scale magnetic field to the required order.  We understood the fact that Faraday's law of electromagnetic induction plays a crucial role in this regard. Therefore, to obtain the correct order of the current magnetic field, understanding the reheating dynamics as well as the evolution of the magnetic field during this period will be of utmost importance. In order to do that, we will consider two possible reheating models and compare the result.

{\bf Case-I:}
 For this we follow the effective one fluid description of reheating dynamics proposed in \cite{Dai:2014jja}, where inflaton energy is assumed to converted into radiation instantaneously at the end of reheating. The dynamics is parametrized by  an effective equation of state $\omega_{eff}$, reheating temperature $T_{re}$ and duration $N_{re}$ (e-folding number during reheating era). In this reheating model, following the approximation as mentioned earlier, one can easily derive the expression for $T_{re}$ and $N_{re}$ in terms of some inflationary parameters as \cite{Cook:2015vqa}
\bea \label{eqtre}
T_{re}= \left(\frac{43}{11 g_{s, re}}\right)^{\frac 1 3}\left(\frac{a_0T_0}{k'}\right) H_k e^{-N_k} e^{-N_{re}}~~,
\eea
  \bea \label{eqnre}
 N_{re}= \frac{4}{(1-3\omega_{eff})} \left [-\frac{1}{4} ln \left(\frac{45}{\pi^2 g_{re}}\right) - \frac{1}{3} ln \left(\frac{11 g_{s,re}}{43}\right)-ln \left(\frac{k'}{a_0 T_0}\right)-ln \left(\frac{V_{end}^{1/4}}{H_k}\right)-N_k \right] ~,
 \eea 
 where the present CMB temperature $T_0=2.725~ K$, the pivot scale ${k'}/{a_0}=0.05$ $M_{pc}^{-1}$. $a_0$ is the present cosmological scale factor. Here for simplicity we have taken both the values of the degrees of freedom for entropy at reheating ($g_{s,re}$), and the effective number of relativistic species upon thermalization ($g_{re}$) is same $g_{s,re}=g_{re}\approx100$. From the above expressions (\ref{eqtre}) and (\ref{eqnre}), we can clearly see that the inflationary parameters put constraints on the reheating parameters $T_{re}$ and $N_{re}$. Considering simple canonical inflation potential $V(\phi)$, the inflation model-dependent input parameters, the inflationary e-folding number $(N_k)$, and the inflationary Hubble constant $(H_k)$  for a particular CMB scale $k$ are known to be written as
 \bea \label{nk}
N_k  = \log\left(\frac{a_{end}}{a_k}\right)= \int_{\phi_{k}}^{\phi_{end}} \frac{|d\phi|}{\sqrt{2\epsilon_v} M_p} ~~,~~H_k =  \frac{\pi M_p\sqrt{r_k A_s}}{\sqrt{2}}~,
\eea
 here the inflaton field $\phi_{end}$ defined at the end of the inflation. Furthermore, the scalar spectral index $n_s^k$, and tensor to scalar ratio $r_k$ can be related with the above inflationary parameters $(H_k, N_k)$ through the following equations
 \begin{equation}
 n_{s}^{k}= 1- 6 \epsilon(\phi_k)+ 2 \eta(\phi_k)~,~r_k=16\epsilon(\phi_k)~~.
\end{equation}
Here the slow-roll parameters expressed as
\begin{equation}
 \epsilon_v= \frac{M_{p}^2}{2}\left(\frac{V'}{V}\right)^2~~;~~|\eta_v|  = M_{p}^2 \frac{|V''|}{V}~.
\end{equation}
The above expressions manage to write reheating parameters in terms of the scalar spectral index ($n_s^k$) for a given CMB scale $k$. After identifying all the required parameters, we will be able to connect CMB and reheating through inflation.\\
{\bf Reheating parameters and primordial magnetic field :} Now in the context of inflationary magnetogenesis, as we mentioned earlier, the electric field continues to exist after the post inflationary era until the universe becomes perfect conductor. This is the non-zero electric field during reheating whose dynamics will significantly change the dynamics of magnetic field and produce the strong magnetic field today. \\ 
For this present reheating model, the phase parameter at the end of the reheating $\theta_k^{re}$ in equation (\ref{eq33}) can be defined as,
 \bea \label{phasek}
 \theta_k^{re}=\left\lbrace Arg\left[\alpha_k\left(z_{end}\right)\beta_k^{*}(z_{end})\right]-\pi\right\rbrace-\frac{4}{3~\omega_{eff}+1}\left(\frac{k}{a_{re}H_{re}}-\frac{k}{a_{end}H_{Inf}}\right)~,
\eea
where $H_{re}$ is the Hubble rate at the end point of reheating. Using evolution of effective density $\rho \propto a^{-3(1+\omega_{eff})}$, one can extract the information about the Hubble parameter at the end of reheating ($H_{re}$) as
\bea
H_{re}=H_{inf} A_{re}^{-\frac{3}{2}(1+\omega_{eff})}~~,
\eea
with the normalized scale factor  $A_{re}={a_{re}}/{a_{end}}=e^{N_{re}}$.
For integer values of $n$, the earlier expression of the phase parameter turns out as
\bea \label{phasekint}
\theta_k^{re}\simeq \frac{2}{2n-1}\frac{k}{a_{end}H_{inf}}\left\{1+\frac{4n-2}{3\omega_{eff}+1}\left(\frac{a_{end}H_{inf}}{a_{re}H_{re}}-1\right)\right\}~~.
\eea
Furthermore,  utilizing equations (\ref{phasekint}) and (\ref{eq32}), one can obtain the magnetic power spectrum during the reheating epoch as,
\beq
\mathcal{P}_{B_{re}}(k)\simeq \frac{8\Gamma(n-\frac{1}{2})^2}{\pi^3H_{Inf}^{-4}}\left(\frac{k}{2a_{end}H_{inf}}\right)^{-2(n-3)}\left(\frac{a_{end}}{a}\right)^4\left\lbrace 1+\left(\frac{4n-2}{3\omega_{eff}+1}\right)\left(\frac{a_{end}H_{inf}}{aH}-1\right)\right\rbrace^2~~.
\eeq
After reheating, the conductivity of the universe becomes sufficiently large. In consequence of that, the electric field dies out very fast, and the magnetic field redshifts as $\mathcal{P}_{B}\propto a^{-4}$ till today.
Since the comoving magnetic power spectrum is conserved after  reheating, the present-day magnetic field obeys the following relation \cite{Kobayashi:2019uqs}
\beq \label{present1}
\mathcal{P}_{B_{0}}(k)\simeq \frac{8\Gamma(n-\frac{1}{2})^2}{\pi^3H_{Inf}^{-4}}\left(\frac{k}{2a_{end}H_{Inf}}\right)^{-2(n-3)}\left(\frac{a_{end}}{a_0}\right)^4\left\lbrace 1+\left(\frac{4n-2}{3\omega_{eff}+1}\right)\left(\frac{a_{end}H_{inf}}{a_{re}H_{re}}-1\right)\right\rbrace^2~~,
\eeq
where the ratio between the scale factor $a_0$ and $a_{re}$, considering entropy conservation can be expressed as 
\bea \label{entrocon}
\frac{a_0}{a_{re}}=\left(\frac{11g_{s,re}}{43}\right)^{\frac{1}{3}} \frac{T_{re}}{T_{0}}~~.
\eea
From the preceding expression, we can see that the magnetic field's present strength explicitly depends on the reheating parameters as well as some inflationary parameters. Therefore,  this opens up the window for probing the early stage of the universe, particularly the reheating phase through the current observational amplitude of the magnetic field. Basically, for a specific value of the spectral index, any present value of the magnetic field $\mathcal{P}_{B_0}$ has a direct one to one correspondence with the effective equation of state of reheating $\omega_{eff}$ and reheating temperature $T_{re}$. Therefore, important conclusion we arrived at that {\em the effective equation of state is no longer a free parameter rather it can be fixed by the present value of $\mathcal{P}_{B_0}$ via CMB}.

{\bf Case-II:} 
For the previous case we did not take into account explicit decay of the inflaton field and additionally the effective equation of state was assumed to be constant. In this case we consider perturbative reheating scenario {\cite{ Maity:2018dgy} constrained by CMB anisotropy. Therefore, as opposed to the previous case, the effective equation of state is time-dependent. However, average inflaton equation of state is taken to be constant. For example inflaton potential $V(\phi) \propto\phi^p$ gives rise to the value of average equation of state
\bea \label{avge}
\omega_\phi =\frac{p-2}{p+2}~~,
\eea
considering virial theorem \cite{mukhanov}. Furthermore, for simplicity we assume  inflaton decays into radiation only. The Boltzmann equation for the inflaton energy density ($\rho_\phi$) and radiation energy density ($\rho_r$) are,
 \bea \label{perturbative 1}
 \Phi^{'} + \frac{\sqrt{3} M_p \Gamma_\phi}{m_{\phi}^2} (1+w_\phi) \frac{A^{1/2}\Phi}{\frac{\Phi}{A^{3 w_\phi}}+ \frac{R}{A}}=0,~~\\\nonumber
  R^{'}- \frac{\sqrt{3} M_p \Gamma_\phi}{m_{\phi}^2} (1+w_\phi) \frac{A^\frac{3 (1-2 w_\phi)}{2}\Phi}{\frac{\Phi}{A^{3 w_\phi}}+ \frac{R}{A}}= 0~~.
 \eea
 Where the comoving densities in terms of the dimensionless variable are used, 
\bea\label{rescale}
\Phi= \frac{\rho_{\phi} A^{3(1+w_\phi)}}{m_\phi^4}  ~ ;  ~ R(t)= \frac{\rho_R A^4}{m_\phi^{4}}~~.
 \eea
 In the above equations to increase the stability of the numerical solution, we use the inflaton mass ($m_\phi$) to define the dimensionless scale factor as $A={a}/{a_{end}}=a m_\phi$.
For solving the Boltzmann equation,  the natural initial conditions will be set at the end of inflation as follows,
.
 \begin{equation}\label{boundary1}
 \varPhi(A=1)= \frac{3}{2} \frac{V_{end}(\phi)}{m_{\phi}^4}~~;~~R(A=1)= 0~.
\end{equation}
For this mechanism the reheating temperature is identified from the radiation temperature $T_{rad}$ at the point of $H(t_{re})=\varGamma_{\phi}$, when maximum inflaton energy density transfer into radiation.
\bea \label{reheating 2}
T_{re}= T_{rad}^{end}= \left(\frac{30}{\pi^2 g_{re}}\right)^{1/4}\rho_{R}(\Gamma_\phi,A_{re},n_{s}^k)^{1/4}~.
\eea
To establish one to one correspondence between $T_{re}$ and $\Gamma_\phi$, we combine the equations (\ref{reheating 2}) and (\ref{eqtre}). To fixed the values of decay width $\Gamma_\phi$ in terms of spectral index ($n_s$), we use one further condition at the end of the reheating
\bea \label{reheating 1}
 H(A_{re})^2= \left(\frac{\dot A_{re}}{A_{re}}\right)^2= \frac{\rho_\phi(\Gamma_\phi,A_{re},n_{s}^k)+ \rho_{R}(\Gamma_\phi,A_{re},n_{s}^k))}{3 M_p^2}=\Gamma_{\phi}^2~~.
\eea 
{\bf Connecting reheating and primordial magnetic field :} In order to connect the reheating and primordial magnetic field through the CMB, it is necessary to understand the cosmological evolution of the electromagnetic field during the post inflationary epoch, especially during reheating, which modifies the present strength of the magnetic power spectrum. As we consider the perturbative decay of the inflaton field during reheating, the phase parameter (\ref{eq33}) now explicitly depends on the evolution of the two energy components, $\rho_\phi$ and $\rho_R$ with time 
\bea \label{phasepert}
\theta_k^{re}= \left\lbrace Arg\left[\alpha_k\left(z_{end}\right)\beta_k^{*}(z_{end})\right]-\pi\right\rbrace-2 k \int\limits_{a_{end}}^{a_{re}}\frac{\sqrt{3} M_p}{\sqrt{\rho_\phi(a)+\rho_R(a)}}\frac{da}{a^2}~.
\eea
Furthermore, for integer values of coupling parameters at the super horizon scale, one can find \cite{Kobayashi:2019uqs}
\bea\label{phaseintper}
\theta_k^{re}\simeq\frac{2}{2n-1}\frac{k}{a_{end}H_{Inf}}\left\lbrace 1+\left(2n-1\right)\int\limits_{a_{end}}^{a_{re}}\frac{\sqrt{3} M_p}{\sqrt{\rho_\phi(a)+\rho_R(a)}}\frac{a_{end}H_{inf}}{a}\frac{da}{a}\right\rbrace~,
\eea
where the time-dependent density components will be followed from Eq.(\ref{perturbative 1}).  Thus, after combining equations (\ref{eq35}), (\ref{entrocon}), and (\ref{phaseintper}), one can obtain the magnetic power spectrum in the present universe as
\bea \label{presentmag}
\mathcal{P}_{B0}(k)\simeq \frac{\Gamma(n-\frac{1}{2})^2}{\pi^3}\frac{2^{2n-3}~(2.6\times 10^{39})}{(6.4\times 10^{-39})^{2n-6}}\left(\frac{k}{a_0}M_{pc}\right)^{-2(n-3)} \left(\frac{11g_{s,re}}{43}\right)^{\frac{2-2n}{3}} \left(\frac{T_{re}}{T_{0}}\right)^{2-2n}\nonumber\\ \left(\frac{H_{inf}}{GeV}~\frac{1}{A_{re}}\right)^{2(n-1)}\left\lbrace 1+\left(2n-1\right)\int\limits_{1}^{A_{re}}\frac{\sqrt{3} M_p H_{Inf}}{\sqrt{\rho_\phi(a)+\rho_R(a)}}\frac{dA}{A^2}\right\rbrace^2 G^2~~.
\eea
In the preceding expression, we can clearly see the appearance of the reheating parameters, which are the function of inflationary observables. Therefore, we definitely will be able to put  indirect bounds on the reheating parameters which in turn can constrain the inflation model. As we emphasized before, from the measurement of the CMB anisotropy, our goal of this paper would be to constraints the reheating dynamics through inflationary parameters considering the present strength of the magnetic field $\mathcal{P}_{B_0}^{{1}/{2}}$.
\section{Magnetic power spectrum in the present universe in terms of reheating parameters: an analytic study}\label{calculation}
Before employing the numerical analysis, in this section we present approximate calculation for the present value of the  magnetic power spectrum and estimate the reheating parameters $(T_{re} ,N_{re})$ in terms that. Following our previous work \cite{Haque:2020zco}, the radiation temperature assumes the following form consider Case-II reheating scenario,
\bea \label{radiation}
T_{rad}\simeq \left\{\frac{2\Gamma_\phi (1+\omega_\phi)}{5-3\omega_\phi}\frac{3M_p^2 H_{Inf}}{\beta A^4}\left(A^{\frac{5-3\omega_\phi}{2}}-1\right)\right\}^{\frac{1}{4}}~~,
\eea
where $\beta={\pi^2 g_{re}}/{30}.
$
Further, utilizing the above expression of the radiation temperature, we can calculate the approximate expression of the reheating temperature. Subsequently, at the point of $A_{re}$, where the condition $H(A_{re})=\Gamma_\phi$ is satisfied, one can define the reheating temperature as
\bea \label{reheatingtemp} 
T_{re}\simeq \left(\frac{2\Gamma_\phi(1+\omega_\phi)}{5-3\omega_\phi}\frac{3M_p^2 H_{inf}}{\beta A_{re}^{\frac{3+3\omega_\phi}{2}}}\right)^{\frac{1}{4}}~~.
\eea
Here the decay width $\Gamma_\phi$ and the normalized scale factor at the end of reheating ($A_{re}$) can be approximated as,
\bea
\Gamma_\phi \simeq \frac{(5-3\omega_\phi) \beta G^4A_{re}^{\frac{3\omega_\phi-5}{2}}}{6M_p^2H_{Inf}}~,~G=\left(\frac{43}{11g_{s,re}}\right)^{\frac{1}{3}}\left(\frac{a_0T_0}{k'}\right)H_k e^{-N_k}~,~A_{re}\simeq \left(\frac{(5-3\omega_\phi)^2 \beta G^4}{12(1+\omega_\phi)^2 M_p^2H_{Inf}^2}\right)^{\frac{1}{1-3\omega_\phi}}~~.
\eea
From the Eq.\ref{eq32} it is obvious that during reheating phase the electromagnetic power spectrum is crucially dependent upon the evolution of the phase $\theta_k$ which is giving rise to induced magnetic field. 
Importantly after the reheating phase ends
large scale magnetic field freezes inside the plasma. Therefore, it is the hierarchy between the inflationary and reheating scale which set the current strength of the magnetic field today after the inflation. Hence naturally one can obtain interesting constraints on reheating evolution parameters such as $(T_{re}, N_{re}, w_{eff})$ though the current value of the large scale magnetic field. As the inflaton energy density dominates large part of the duration of the  reheating phase, we approximate the solution of the inflaton energy density as 
\bea
\rho_\phi(A)=\rho_\phi^{end}A^{-3(1+\omega_\phi)}e^{-\Gamma_\phi(t-t_i)} \simeq \rho_\phi^{end}A^{-3(1+\omega_\phi)}~~.
\eea
This helps us to further obtaining the approximate Hubble parameter as
\begin{eqnarray}
H(A)=\sqrt{\frac{\rho_\phi(A)+\rho_R(A)}{3M_p^2}}\simeq \frac{H_{Inf}}{A^{\frac{3(1+\omega_\phi)}{2}}}\sqrt{e^{-\Gamma_\phi(t-t_i)}+\frac{A^{\frac{3(1+\omega_\phi)}{2}}}{\eta}}\simeq \frac{H_{Inf}}{A^{\frac{3(1+\omega_\phi)}{2}}}\sqrt{1+\frac{A^{\frac{3(1+\omega_\phi)}{2}}}{\eta}}~~,
\end{eqnarray}
where $\eta=\frac{(5-3\omega_\phi)H_{inf}}{2\Gamma_\phi(1+\omega_\phi)}$. With all the above approximate expressions, 
one can arrive the following expression of the present magnetic power spectrum
\begin{equation}
\begin{split} \label{presentmaganalytical}
\mathcal{P}_{B0}(k)\simeq \frac{\Gamma(n-\frac{1}{2})^2}{\pi^3}\frac{2^{2n-3}~(2.6\times 10^{39})}{(6.4\times 10^{-39})^{2n-6}}\left(\frac{k}{a_0}M_{pc}\right)^{-2(n-3)} \left(\frac{11g_{s,re}}{43}\right)^{\frac{2-2n}{3}} \left(\frac{T_{re}}{T_{0}}\right)^{2-2n}\left(\frac{H_{inf}}{GeV}~\frac{1}{A_{re}}\right)^{2(n-1)} \\ \left\{ 1+\left(2n-1\right)\left[\frac{2}{3\omega_\phi+1}\left(\eta^{\frac{3\omega_\phi+1}{3(1+\omega_\phi)}}-1\right)+\frac{4}{3\omega_\phi-1}\eta^{1/2}\left(A_{re}^{\frac{3\omega_\phi-1}{4}}-\eta^{\frac{3\omega_\phi-1}{6(1+\omega_\phi)}}\right)\right]\right\}^2 G^2~.
\end{split}
\end{equation}
Interestingly, our analytical expression of the magnetic power spectrum roughly matches with the numerical values.
\section{Inflation models and numerical results}\label{sec3}
  The main aim of our study is to see how the present observational limit on the magnitude of large scale magnetic field as well as the CMB anisotropy can be used to probe the reheating dynamics and put  combined constraints on the reheating and inflationary parameter space.
We consider different inflationary models ans study how the present limits on the magnetic field strength constrain the effective reheating equation of state $\omega_{eff}$ in terms of inflationary scalar spectral index $n_s$. Further, the allowed range of the current magnetic field can be shown to impose an upper limit on the values of the spectral index $n_s^{max}$ and effective equation of state $\omega_{eff}^{max}$, which in turn provides  bound on the maximum possible reheating temperature $T_{re}^{max}$. Moreover, to connect reheating parameters such as $(T_{re}$, $\omega_{eff})$ with the present magnetic field power spectrum $\mathcal{P}_{B_0}^{1/2}$, we take into account the gauge kinetic function with power $n=2$. We also studied interesting constraints on the maximum possible value of the magnetogenesis model parameter   $n=n_{max}$ and associated maximum  reheating temperature $T_{re}^{max}$ for different inflationary model.
\subsection{Geneal discussion on our  results}
 In the conventional study the magnetic energy density is assumed to be diluted adiabatically with the expansion of the universe as $1/a^4$ starting from the end of inflation. This framework never gives rise to enough present day magnetic field strength for high reheating temperature, which is believed to be generic prediction of inflation in the effective field theory framework. Therefore, in the inflationary magnetogenesis scenario non-trivial magnetic field dynamics during reheating should be an important physical phenomena that has to be  understood quite well. Furthermore, during this phase itself the inflaton decay should happen in such a manner that the electric field survives for Faraday's effect to play the role. 

Most important point of our study is the constraints set on the effective reheating  equation of state $\omega_{eff}$. A particular present day value of $P_{B_0}$, the CMB anisotropic constraint and after reheating entropy conservation law  automatically fixes the reheating equation of state uniquely. Most interestingly, this would place severe restriction on the inflaton potential which we will study in our later paper. In our discussions, we mianly concentrate on the scale invariant electric power spectra ($n=2$) for two possible reheating scenarios. One corresponds to instantaneous conversion of inflaton energy to radiation at the end of the reheating. The second one is the  perturbative reheating where inflaton energy density transferred into radiation gradually with finite inflaton decay width.

{\bf{Case-I, Instantaneous reheating model :}} Even though we are considering scale invariant electric field, important to remember that it will survive only until the end of reheating. The generated large scale magnetic field will be constrained by this condition. In this case how the reheating and inflationary parameters and the large scale magnetic field  are intertwined each other are clearly depicted in fig.(\ref{n2}). From the first  panel of fig.(\ref{n2}), we can clearly predict a unique value of $\omega_{eff}$ associated with a specific choice of the present magnetic field once we fixed scalar spectral index $n_s$. Additionally, for a given $\omega_{eff}$, the reheating temperature is also determined uniquely. {\em Therefore, taking into account CMB constraints the reheating phase can be uniquely probed by the evolution of the primordial magnetic field}.\\
In order to have quantitative understanding of our analysis, in the Tables- \ref{axiontab1}, \ref{alphatab1}, \ref{minimaltab1}, important reheating parameters such as $\omega_{eff}$ and $T_{re}$ for different sample values of the current magnetic field are given for a particular values of the spectral index $n_s^{central}=0.9649$.
 Furthermore, for any specific choice of the current magnetic field strength within the observational limit, there exits a maximum allowed values of the spectral index $n_s^{max}$ and associated $\omega_{eff}^{max}$, and this naturally leads to the maximum permissible value of the reheating temperature shown in the tables \ref{axiontab2}, \ref{alphatab2}, \ref{minimaltab2}. We consider three distinct  values of $\mathcal{P}_{B_0}^{1/2}=(10^{-18},10^{-20},10^{-22})$ G through out the discussion. Most importantly results for the case of instantaneous reheating scenario is the  constraints on the effective reheating equation of state $ 0.15 < \omega_{eff} <0.33$ for qudratic inflaton potential  near its minimum. Furthermore, the minimum limiting value of magnetic field ($\textit{P}_{B_0}^{1/2} < 10^{-22}$)$  $ G, set an associated maximum reheating temperature around $\sim$ 1 TeV. Therefore, magnetogenesis models with scale invariant electric field appears to be compatible with only low scale inflationary model.  \\
{\bf{Perturbative reheating model (case-II) :}} The present day magnetic field and the reheating temperature are intimately connected through the inflaton equation of state which can be guessed from the previous analysis. However, perturbative reheating is observed to be better suited in understanding the nature of both inflaton and magnetogenesis model itself. Important characteristic outcome in considering perturbative reheating scenario is that large scale magnetic field permits us to understand the nature of the inflaton field and its observational viability through the constraints on the inflaton equation of state.   
For example, as long as scale invariant power spectra model is concerned, $\omega_{\phi}=0$ can not produce required strength of the magnetic field, which is indicating the fact that inlaton field with quadratic potential near it minimum is not compatible with observation. However, assuming  higher inflaton equation of state $\omega_\phi = (p-2)/(p+2) >\frac{1}{3}$ the observable magnetic field strength can be successfully generated within the perturbative reheating framework shown in the last two plots of  the figs.(\ref{perturbativedifferentw}).
  Further, considering the  magnetic field strength within the observational limit, the maximum and minimum allowed values of the spectral index $n_s$, the inflaton equation of state $\omega_{\phi}$ and the associated reheating temperature are provided in the tables \ref{perturbound2}, \ref{perturbativebound} for different inflationary models. \\
With this general discussion, in the following sections we consider various inflationary models and discuss about their quantitative predictions.

  \begin{figure}[t!] 
 	\begin{center}
 		
 \includegraphics[width=005.55cm,height=03.6cm]{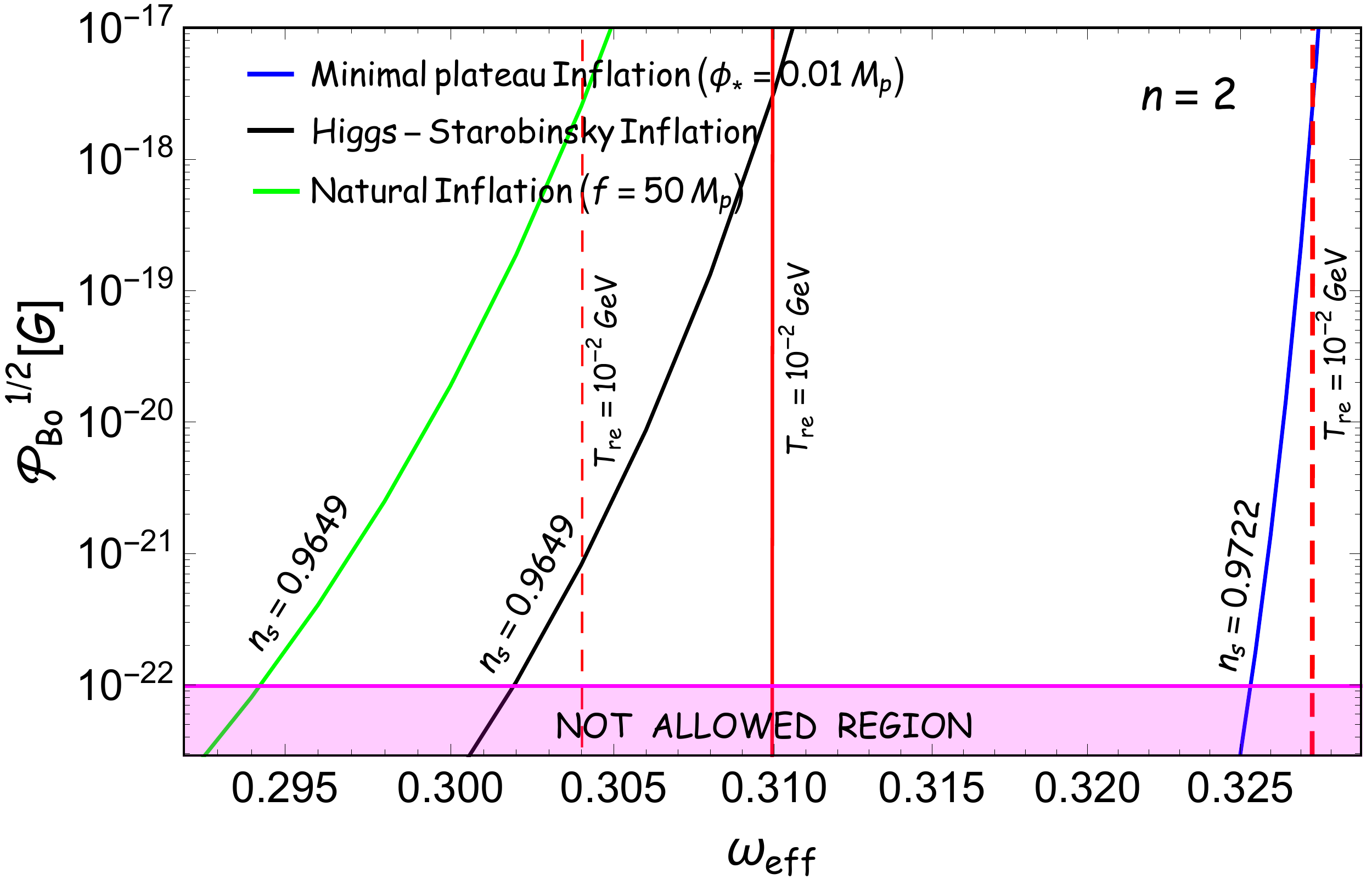}
   \includegraphics[width=5.55cm,height=3.6cm]{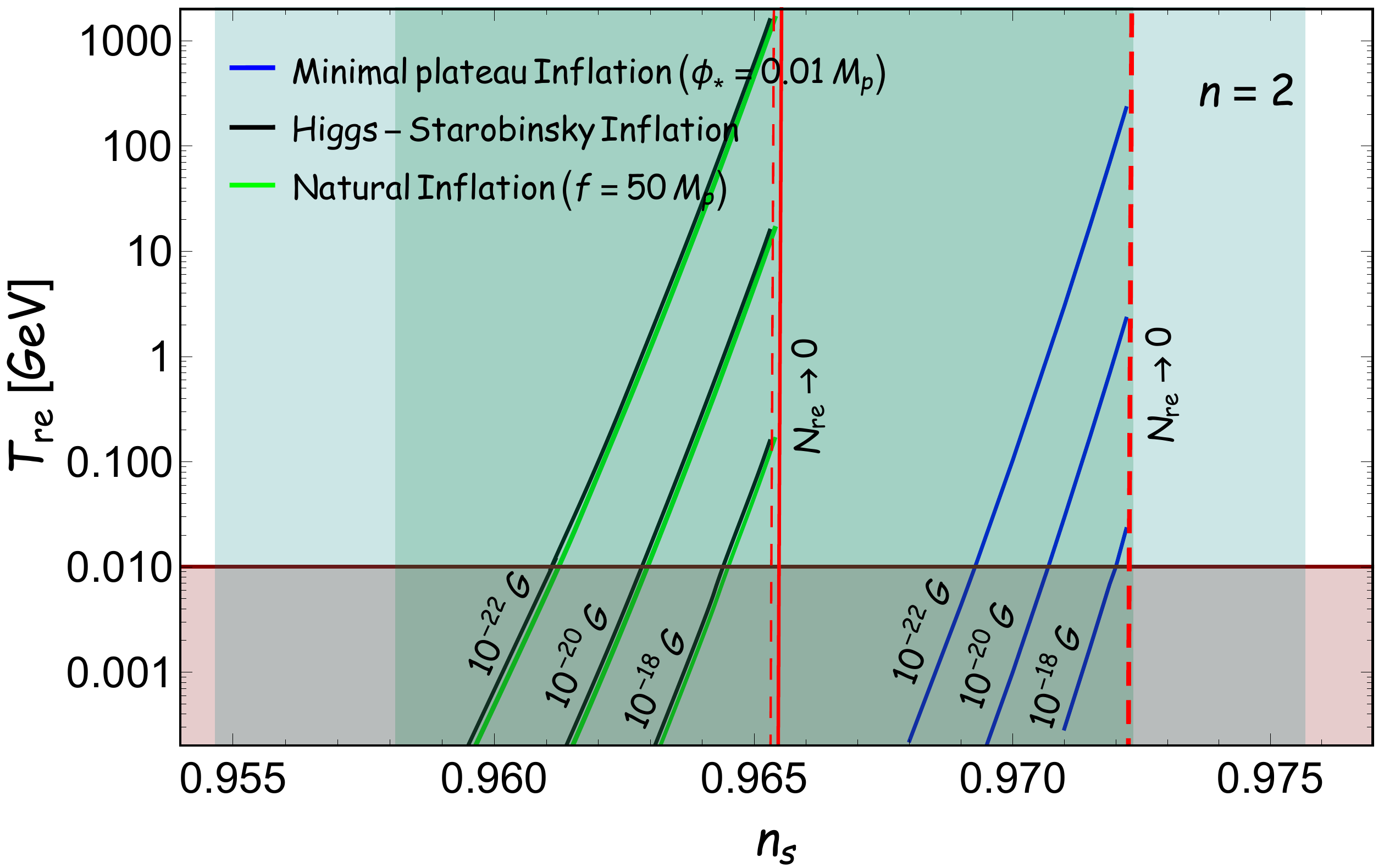}			
  \includegraphics[width=5.55cm,height=3.6cm]{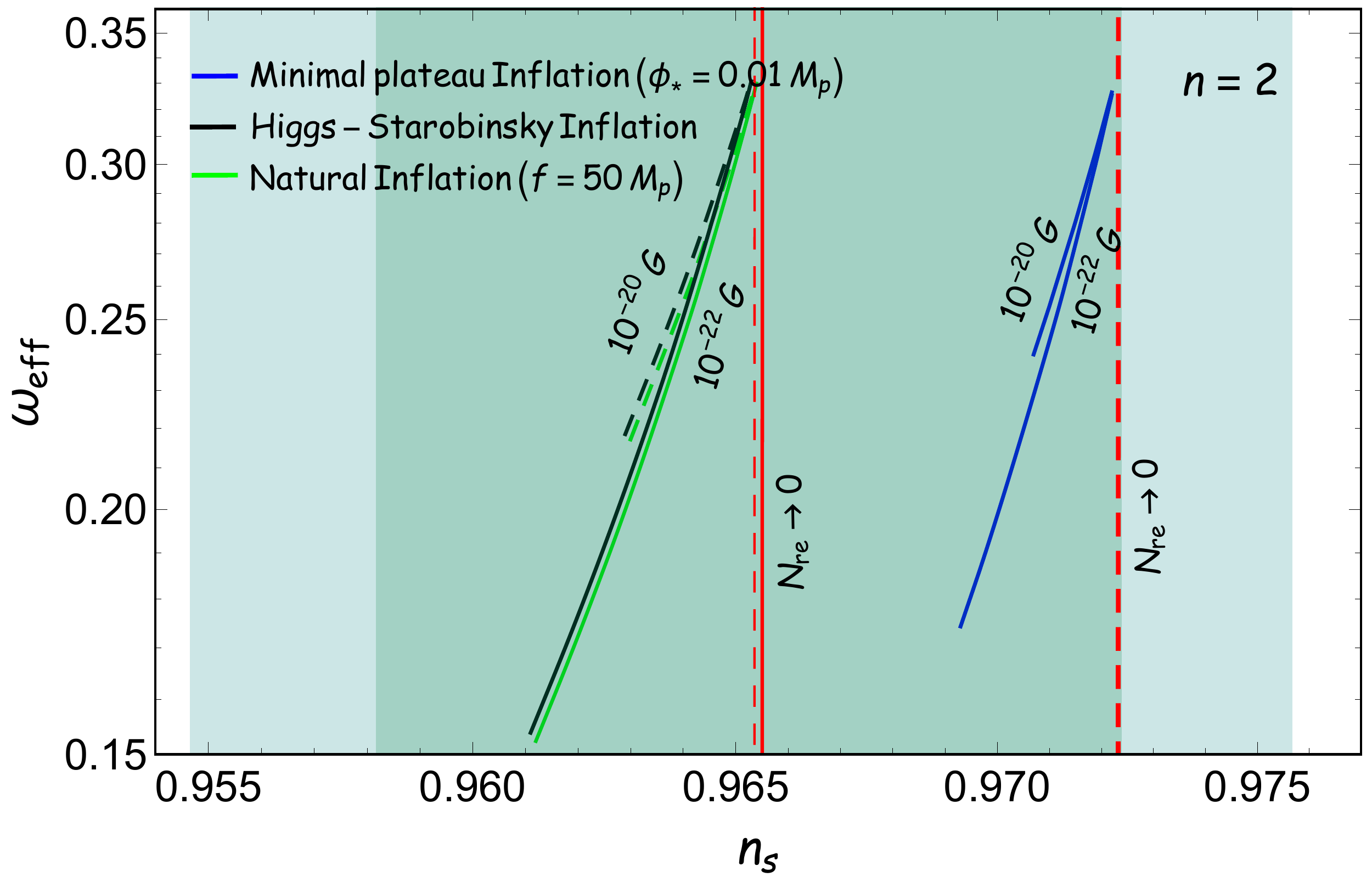}
	\caption{\scriptsize  We plot the magnetic field variation on the upper left side as a function of the effective equation of state for a fixed value of the spectral index. The light pink region indicates the not allowed region as the strength of the magnetic field in that region is less than $10^{-22}$ $G$ ( observed value of the magnetic field on scales $\approx 1 Mpc$ must be $B^0_{cosmic}>10^{-22}$ $G$ \cite{Martin:2007ue}). The allowed value of the present magnetic field range from $10^{-22}$ $G$ to the point where the reheating temperature reaches $10^{-2}$ $GeV$ (comes from BBN constraints), marked as solid and dashed red line. On the middle, we have plotted the variation of reheating temperature as a function of the spectral index for three different values of the current magnetic field $\mathcal{P}_{B_0}^{1/2}=(10^{-18}, 10^{-20},10^{-22})$ G. The light brown region indicates the reheating temperature ($T_{re}$) below $10^{-2}$ $GeV$, which would ruin the predictions of big bang nucleosynthesis (BBN). In the last plot, we have shown the variation of the effective equation of state as a function of the spectral index ($n_s$) with two different values of $\mathcal{P}_{B_0}^{1/2}=(10^{-20},10^{-22})$ G, within the minimum and maximum values of reheating temperature.  The deep and light green band indicates $1\sigma$ and $2\sigma$ range of $n_s$ from Planck \cite{Akrami:2018odb}. Here we have plotted the results for three different inflationary models (Natural inflation, Higgs-Starobinsky inflation, minimal plateau model) for the $k$-independent electric power spectrum ($n=2$). }
 		\label{n2}
 	\end{center}
 \end{figure}

 \subsection{Natural inflation \cite{axion}}
 

 In natural inflationary model, the potential is defined by
\beq
\label{eq39}
V(\phi)=\Lambda^4\left[1-\cos\left(\frac{\phi}{f}\right)\right]~~,
\eeq
 here $\Lambda$ is the height of the potential, and $f$ is the width, also known as the axion decay width. To fit this model with Planck data, this model needs a super-Planckian value of axion decay constant. This is why we have taken $f=50M_p$ for our numerical analysis purpose. Here the overall scale of the inflation $\Lambda$ fixes by the CMB normalization.\\
  In order to connect the primordial magnetic field with those observed in the present universe through reheating dynamics, we need to define inflationary e-folding number $N_k$ and tensor-to -scaler ratio $r_k$. In the framework of the natural inflation model, the inflationary parameters $N_k$ and $r_k$ in terms of $n_s$ are defined as
 \begin{equation}
 N_k=\frac{f^2}{M_p^2} \ln \left(\frac{2f^2 (f^2 (1-n_s)+M_p^2)}{(2f^2 +M_p^2) (f^2 (1-n_s)- M_p^2)}\right)~,~r_k= 4\left(\frac{f^2 (1-n_s)- M_p^2}{f^2}\right)~.
\end{equation}

In addition with that for the perturbative reheating model, the initial conditions to solve the Boltzmann equations for two density component are set at the end of the inflation to be
\begin{equation}
 \varPhi(A=1)=\frac{3}{2} \frac{2 \Lambda^4 M_p^2}{(2 f^2 +M_p^2) m_\phi^4}~,~R(A=1)=\frac{3\left(\omega_{eff}-\omega_\phi\right)}{1-3\omega_{eff}}~\varPhi(A=1)~,
\end{equation}
where
\begin{equation}
 \Lambda=\left(\frac{3\pi^2 M_p^2 A_s (f^4(1-n_s)^2 -M_p^4)}{2 f^2}\right)^{\frac{1}{4}}~,~m_\phi=\frac{\Lambda^2}{f}~.
\end{equation} 
Here, the primordial scalar amplitude of the inflationary scalar fluctuation is $A_{\delta\phi}\sim 2.19\times 10^{-9}$ from Planck \cite{Akrami:2018odb}. 
 As we have the connection relation (Eqs. \ref{present1}, \ref{presentmag}) between the reheating parameters and the parameters of magnetogenesis model, in the following, we discuss their implications and various constraints for inflationary magnetogenesis model with $n=2$ for two distinct reheating scenarios, which we have discussed before.\\
  \begin{figure}[t!]
 	\begin{center}
 \includegraphics[width=008.30cm,height=5.25cm]{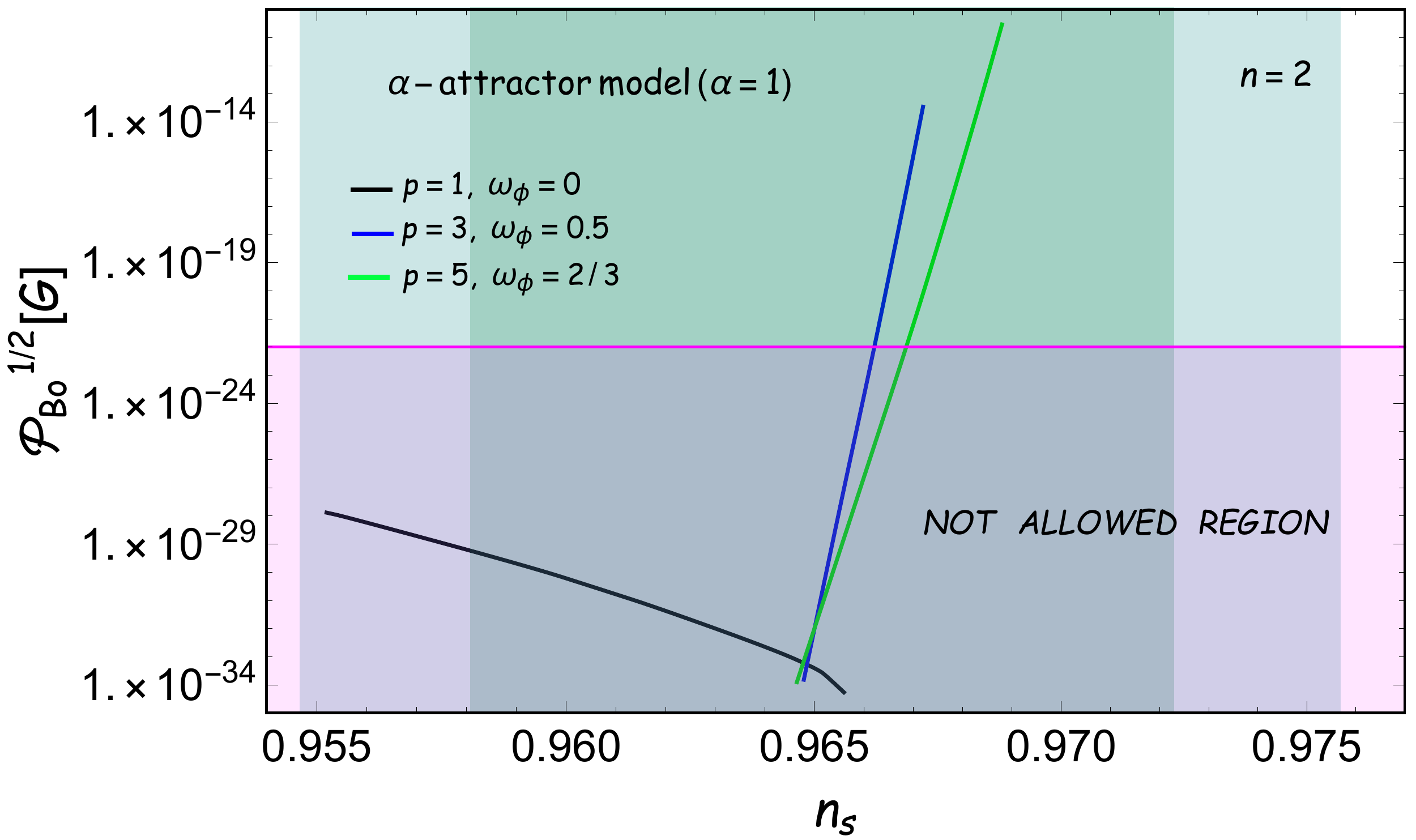}
 \includegraphics[width=008.30cm,height=5.25cm]{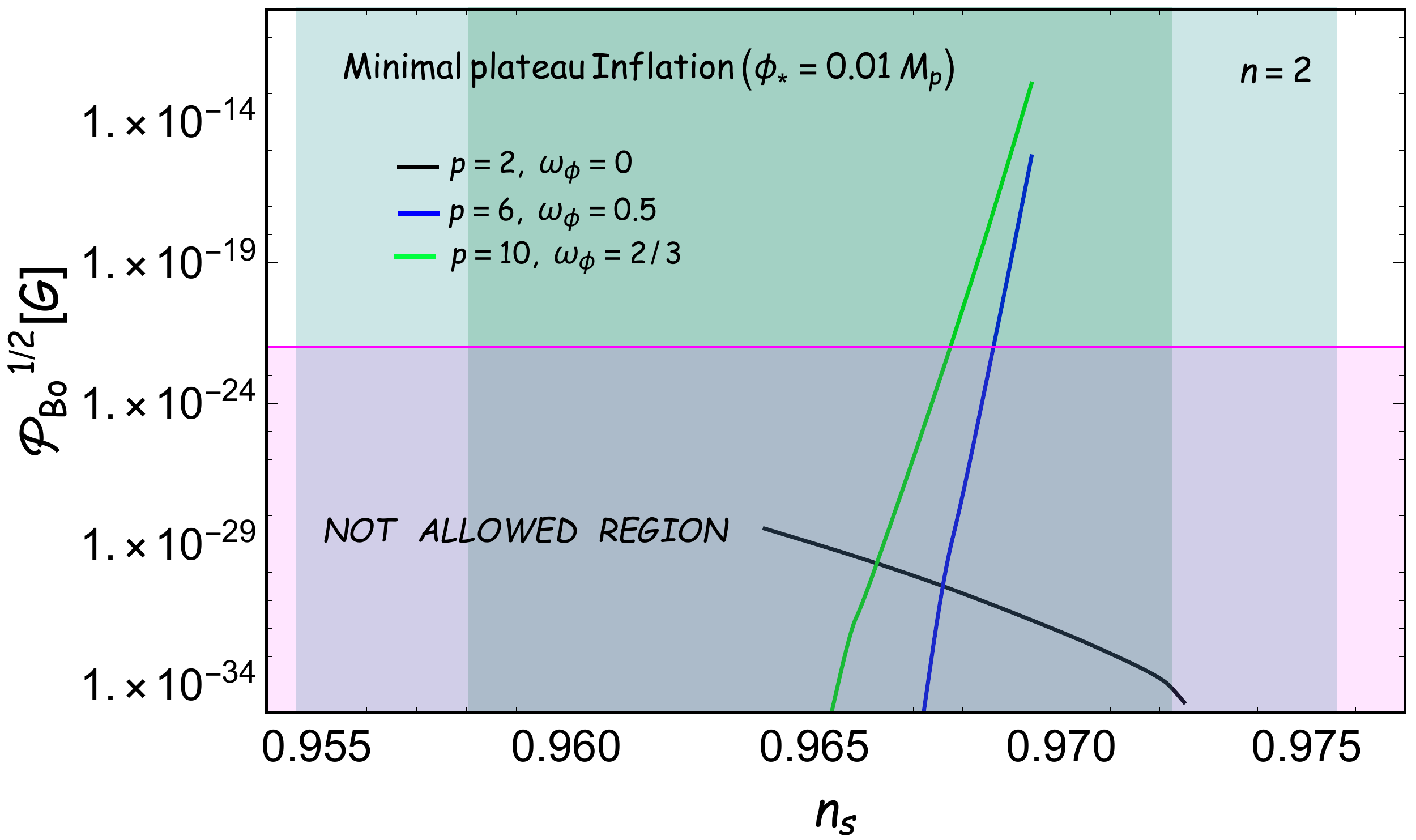}		
\includegraphics[width=008.30cm,height=5.25cm]{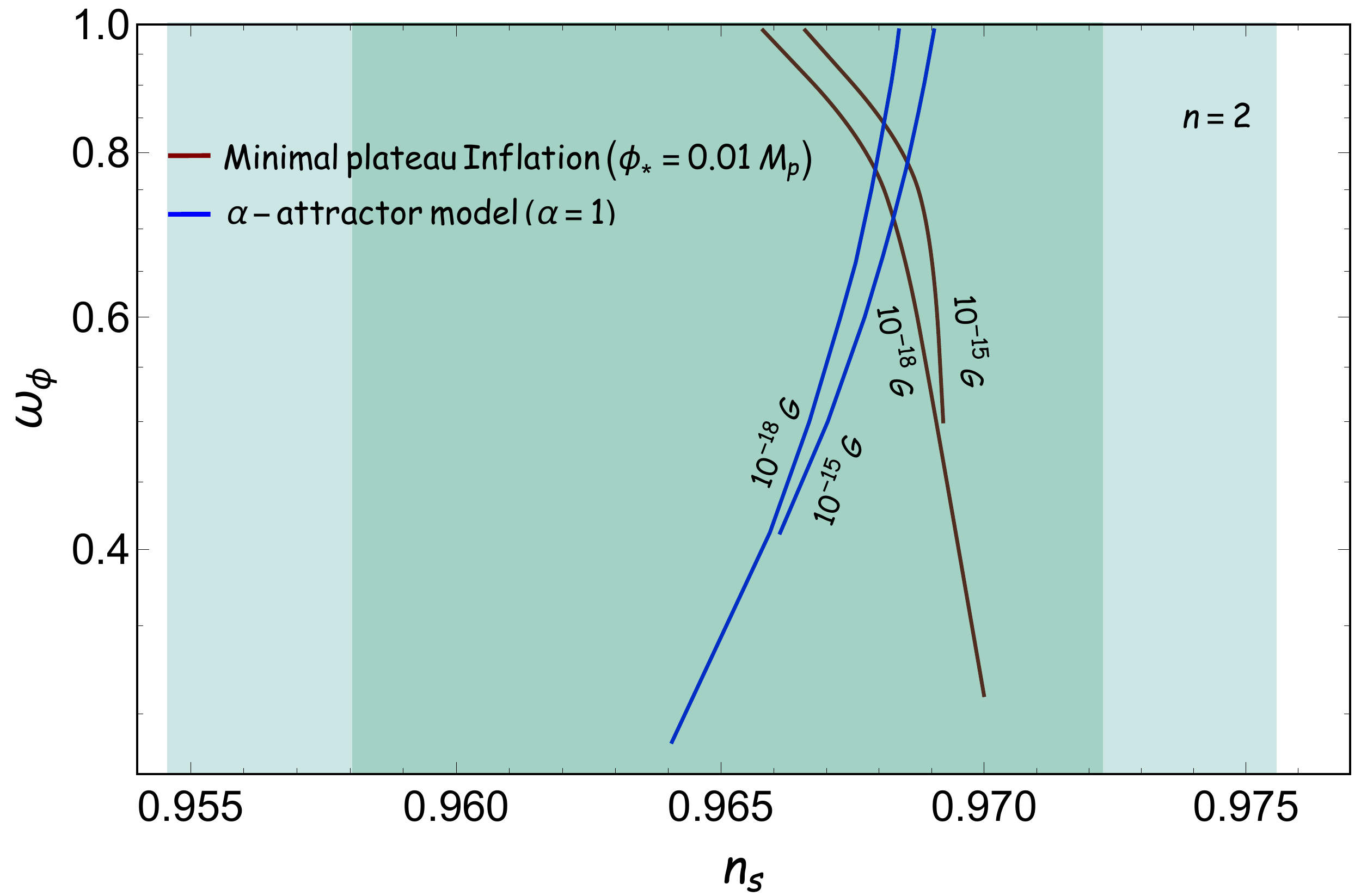}	
\includegraphics[width=008.30cm,height=5.25cm]{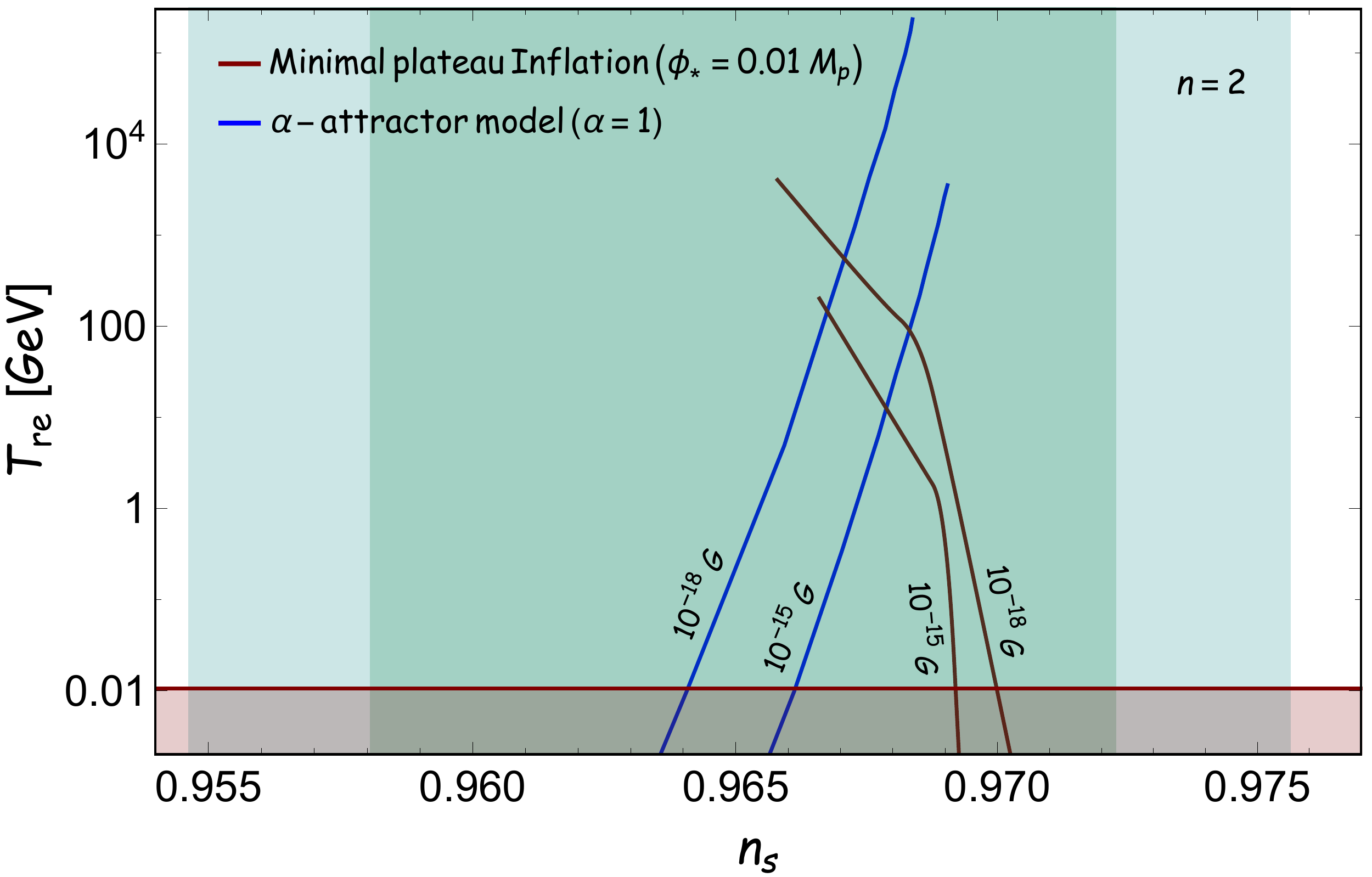}	
		\caption{In the first two plots, we have shown the variation of the present-day magnetic field as a function of the spectral index within the minimum and maximum values of the reheating temperature for different inflationary models, considering the perturbative dynamics of reheating with inflaton equation of state ($\omega_\phi=(0,\frac{1}{2}, \frac{2}{3})$). The light pink region indicates the not allowed region as the strength of the magnetic field in that region is less than $10^{-22}$ $G$ ( observed value of the magnetic field on scales $\approx 1 Mpc$ must be $B^0_{cosmic}>10^{-22}$ $G$ \cite{Martin:2007ue}). We plot on the lower left side the variation of the inflaton equation of state $\omega_\phi$ as a function of the spectral index for two different values of the magnetic field $\mathcal{P}_{B_0}^{1/2}=(10^{-15}, 10^{-18})$ G. On the lower right side, we have plotted the variation of reheating temperature as a function of the spectral index with the same fixed values of the $\mathcal{P}_{B_0}^{1/2}$.  The light brown region indicates the reheating temperature ($T_{re}$) below $10^{-2}$ $GeV$, which would ruin the predictions of big bang nucleosynthesis (BBN). }
 		\label{perturbativedifferentw}
 	\end{center}
 \end{figure}
In the first panel of the Fig.\ref{n2}, for PLANCK central value of $n_s= 0.9649$, the reheating equation of state $\omega_{eff}$ has to be bounded within a very small window $(0.294, 0.304)$ which is close to the radiation equation of state.  The upper bound is associated with the BBN limit of $T_{re} \sim 10^{-2}$ GeV. 
From the third panel of Fig.\ref{n2}, one observes that with the decreasing magnetic field, maximum allowed reheating temperature is increasing which seems to be intuitively obvious as reheating e-folding is accordingly decreasing. All the above result are for the case-I when the reheating is governed by an effective equation of state. 

 For the pertubrvative reheating (case-II) model with $\omega_\phi=0$, the maximum strength of the large scale magnetic field is $\sim 10^{-28}$ G, which is much smaller compared to the observational limit. Within the perturbative reheatinbg framework, the axion inflation model is not compatible with the larg scale magnetic field observation.\\
   \begin{table}[t!]
  
\caption{Probing reheating phase (fixing effective equation of state and reheating temperature)}
Natural inflation model ($f=50M_p$)\\[.1cm]
\begin{tabular}{|p{2.7cm}|p{2.7cm}|p{2.7cm}|p{2.7cm}|p{2.7cm}|}
\hline
 \quad Parameters&\multicolumn{4}{c|}{Current observed value of the magnetic field ($\mathcal{P}_{B_0}^{1/2}$) measured in unit of Gauss}\\
 \cline{2-5}
&\multicolumn{1}{c|}{$n_s$} &\multicolumn{3}{c|}{Scale invariant electric field $(n=2)$}\\
 \cline{3-5}
 &&\qquad$10^{-18}$ G&\qquad$10^{-20}$ G &\qquad$10^{-22}$ G \\
 \hline
 \qquad $\omega_{eff}$&\qquad$0.9649$&\qquad$0.303$&\qquad $0.299$&\qquad $0.294$\\
  \qquad$T_{re}$ (GeV)&&\qquad$0.033$&\qquad $3.36$&\qquad $339.30$\\
\hline 
 \end{tabular}
 \label{axiontab1}
 \end{table}

 \begin{table}[t!]
\caption{
Constraining reheating and inflationary parameters through inflationary magnetogenesis}
Natural inflation model ($f=50 M_p$)\\[.1cm]
\begin{tabular}{|p{3cm}|p{4cm}|p{4cm}|p{4cm}|  }
\hline
 \qquad Parameters&\multicolumn{3}{c|}{Current observed value of the magnetic field ($\mathcal{P}_{B_0}^{1/2}$) measured in unit of Gauss}\\
 \hline
 &\multicolumn{3}{c|}{Scale invariant electric field $(n=2)$}\\
 \cline{2-4}
 &\qquad$10^{-18}$ G&\qquad$10^{-20}$ G &\qquad$10^{-22}$ G\\
\hline
 \qquad $n_s^{min}$&\qquad$0.9645$&\qquad $0.9630$&\qquad $0.9612$\\
 \qquad $n_s^{max}$&\qquad$0.9654$&\qquad $0.9654$&\qquad $0.9654$\\
   \qquad$\omega_{eff}^{min}$&\qquad$0.2842$&\qquad $0.2171$&\qquad $0.1523$\\
   \qquad $\omega_{eff}^{max}$&\qquad$0.3307$&\qquad $0.3303$&\qquad $0.3298$\\
    \qquad $T_{re}^{min}$(GeV)&\qquad$10^{-2}$&\qquad $10^{-2}$&\qquad$10^{-2}$\\
   \qquad $T_{re}^{max}$(GeV)&\qquad$0.16$&\qquad $16.42$&\qquad$1.6\times 10^3$\\
    \hline
 \end{tabular}
 \label{axiontab2}
 \end{table} 
   
    \subsection{\bf $\alpha-$attractor model \cite{alpha}}
    
 		
 		
 
   \begin{table}[t!]
  
\caption{Probing reheating phase (fixing effective equation of state and reheating temperature)}
Higgs-Starobinsky inflation model\\[.1cm]
\begin{tabular}{|p{2.7cm}|p{2.7cm}|p{2.7cm}|p{2.7cm}|p{2.7cm}|}
\hline
 \quad Parameters&\multicolumn{4}{c|}{Current observed value of the magnetic field ($\mathcal{P}_{B_0}^{1/2}$) measured in unit of Gauss}\\
 \cline{2-5}
&\multicolumn{1}{c|}{$n_s$} &\multicolumn{3}{c|}{Scale invariant electric field $(n=2)$}\\
 \cline{3-5}
 &&\quad$10^{-18}$ G&\quad$10^{-20}$ G &\quad$10^{-22}$ G \\
 \hline
  $\omega_{eff}$&\qquad$0.9649$&\quad$0.3093$&\quad $0.3061$&\quad $0.3019$\\
  $T_{re}$ (GeV)&&\quad$0.044$&\quad $4.45$&\quad $447.62$\\
\hline 
 \end{tabular}
 \label{alphatab1}
 \end{table} 

\begin{table}[t!]
\caption{
Constraining reheating and inflationary parameters through inflationary magnetogenesis}
Higgs-Starobinsky inflation model\\[.1cm]
\begin{tabular}{|p{3cm}|p{4cm}|p{4cm}|p{4cm}|  }
\hline
 \qquad Parameters&\multicolumn{3}{c|}{Current observed value of the magnetic field ($\mathcal{P}_{B_0}^{1/2}$) measured in unit of Gauss}\\
 \hline
 &\multicolumn{3}{c|}{Scale invariant electric field $(n=2)$}\\
 \cline{2-4}
 &\qquad$10^{-18}$ G&\qquad$10^{-20}$ G &\qquad$10^{-22}$ G\\
\hline
 \qquad$n_s^{min}$&\qquad$0.9644$&\qquad $0.9629$&\qquad $0.9611$\\
  \qquad$n_s^{max}$&\qquad$0.9653$&\qquad $0.9653$&\qquad $0.9653$\\
  \qquad $\omega_{eff}^{min}$&\qquad$0.2852$&\qquad $0.2185$&\qquad $0.1538$\\
   \qquad $\omega_{eff}^{max}$&\qquad$0.3314$&\qquad $0.3311$&\qquad $0.3307$\\
    \qquad $T_{re}^{min}$(GeV)&\qquad$10^{-2}$&\qquad $10^{-2}$&\qquad$10^{-2}$\\
    \qquad $T_{re}^{max}$(GeV)&\qquad$0.16$&\qquad $15.7$&\qquad$1.6\times 10^3$\\
    \hline
 \end{tabular}
 \label{alphatab2}
 \end{table} 
 
   \begin{table}[t!]
\caption{
Constraining reheating and inflationary parameters through inflationary magnetogenesis (Perturbative reheating dynamics)}
\begin{tabular}{|p{3cm}|p{3cm}|p{3cm}|p{3cm}|p{3cm}| }
\hline
 \qquad Parameters&\multicolumn{4}{c|}{Current observed value of the magnetic field ($\mathcal{P}_{B_0}^{1/2}$) measured in unit of Gauss}\\
 \cline{2-5}
 &\multicolumn{4}{c|}{Scale invariant electric field $(n=2)$}\\
 \cline{2-5}
 &\multicolumn{2}{c|}{$10^{-15}$ G}&\multicolumn{2}{c|}{$10^{-18}$ G} \\
 \cline{2-5}
&\multicolumn{1}{c|}{$\alpha-$attractor ($\alpha=1$)}&\multicolumn{1}{c|} {minimal~($\phi_*=0.01M_p)$}&\multicolumn{1}{c|}{$\alpha-$attractor ($\alpha=1$)}&\multicolumn{1}{c|} {minimal~($\phi_*=0.01M_p)$}\\
\hline
 \qquad$n_s^{min}$&\qquad$0.966125$&\qquad $0.969224$&\qquad $0.964075$&\qquad $0.96580$\\
  \qquad$n_s^{max}$&\qquad$0.96905$&\qquad $0.9666$&\qquad $0.96838$&\qquad$0.9700$\\
  \qquad $\omega_{\phi}^{min}$&\qquad$0.412$&\qquad $0.500$&\qquad $0.286$&\qquad$0.3100$\\
   \qquad $\omega_{\phi}^{max}$&\qquad$0.99$&\qquad $0.99$&\qquad $0.99$&\qquad$0.99$\\
    \qquad $T_{re}^{min}$(GeV)&\qquad$10^{-2}$&\qquad $10^{-2}$&\qquad$10^{-2}$&\qquad$10^{-2}$\\
    \qquad $T_{re}^{max}$(GeV)&\qquad$3.5\times 10^3$&\qquad $200.1$&\qquad$2.32\times 10^5$&\qquad$4.0\times 10^3
    $\\
    \hline
 \end{tabular}
 \label{perturbound2}
 \end{table} 
In this section, we will consider the $\alpha-$attractor model, which is a class of theoretical models that unifies a large number of the existing inflationary models proposed in \cite{alpha}. Conformal transformation of a large class of non-canonical inflaton field lagrangian, a canonical exponential potential obtained in the following form
 \begin{equation} \label{a}
V(\phi) = \Lambda^4 \left[  1 - e^{ -     \sqrt{\frac{2}{3\alpha} }    \frac{\phi}{M_p}     } \right]^{2p}.
\end{equation}
The canonical property of this class of model predicts inflationary observable $(n_s,r)$, compatible with Planck observation \cite{Ade:2015lrj,Akrami:2018odb}. Here the mass scale $\Lambda$ of the E-model fixed from CMB normalization. However, this class of models includes the Higgs-Starobinsky model \cite{Bezrukov:2007ep,Starobinsky:1980te} for $p=1$, $\alpha=1$.\\ 
Before we jump into our analysis details, let us first determine the relationship between the inflationary parameters with potential parameters. The inflationary e-folding number, $N_k$ and tensor to scalar ratio, $r_k$, can be express interms of $n_s$ and model parameter as
 \bea
  N_k=\frac{3\alpha}{4p} \left[e^{\sqrt{\frac{2}{3\alpha}}\frac{\varPhi_k}{M_p}}-e^{\sqrt{\frac{2}{3\alpha}}\frac{\varPhi_{end}}{M_p}}-\sqrt{\frac{2}{3\alpha}}\frac{(\varPhi_k-\varPhi_{end})}{M_p}\right]~,~r_k=\frac{64 p^2}{3\alpha \left(e^{\sqrt{\frac{2}{3\alpha}}\frac{\varPhi_k}{M_p}}-1\right)^2}~~.
\eea
In addition to that, the initial conditions  to solve the Boltzmann equations for different energy components, considering the perturbative reheating model in the context of the present scenario, can be expressed as
\bea
 \varPhi(A=1)= \frac{3}{2} \Lambda^4 \left(\frac{2p}{2p+\sqrt{3 \alpha}}\right)^{2p}~~,~~R(A=1)=0~,
 \eea
 where 
 \bea
 \Lambda= M_p \left(\frac{3\pi^2 r A_s}{2}\right) \left[\frac{2p(1+2p)+\sqrt{4p^2+6\alpha(1+p)(1-n_s)}}{4p(1+p)}\right]^{\frac{p}{2}}~.
 \eea
We consider $\alpha-$attractor model with $\alpha=1$ for two different reheating scenarios  described before. Detail prediction and model constraints can be read from Figs.(\ref{n2}), (\ref{perturbativedifferentw}) and tables \ref{alphatab1}, \ref{alphatab2}, \ref{perturbativebound}. Taking a particular central value of the scalar spectral, $n_s=0.9649$, the effective equation of state is constrained  within a very narrow range $(0.302,0.310)$. As has been mentioned already and also seen for the axion model, large scale magnetic field appeared to be consistent with low reheating temperature with maximum value $\sim$ 1 TeV for $\mathcal{P}_{B_0}^{1/2}\sim 10^{-22}$ G. The above conclusion are based on instantaneous reheating scenario. \\ 
For perturbative case, $\omega_\phi=0$ turned out to be observationally unsuitable. However, for the model with $p=(3,5)$ where the inflaton equation states assume $\omega_\phi=(0.5,\frac{2}{3})$ accordingly, are observed to be potentially consistent with observation. Most importantly the perturbative reheating supports higher value of reheating temperature $T_{re}^{max}\sim(2.4\times 10^4, 2.1\times 10^6)$ GeV for magnetic field strength $\mathcal{P}_{B_0}^{\frac{1}{2}}\sim 10^{-22}G$, with $\omega_\phi=(0.5,\frac{2}{3})$ respectively. Quantitative values of all the reheating parameters and inflationary parameters $(n_s,\omega_\phi,T_{re})$ corresponding to different sample values of $\mathcal{P}_{B_0}^{1/2}$ are provided in table-\ref{perturbound2}. This can provide us clear picture of the viability of the $\alpha-$attractor model in the context of inflationary magnetogenesis scenarios.

  \subsection{Minimal plateau model}
 \begin{table}[t!]
  
\caption{Probing reheating phase (fixing effective equation of state and reheating temperature)}
Minimal plateau model ($\phi_*=0.01M_p$)\\[.1cm]
\begin{tabular}{|p{2.7cm}|p{2.7cm}|p{2.7cm}|p{2.7cm}|p{2.7cm}|}
\hline
 \quad Parameters&\multicolumn{4}{c|}{Current observed value of the magnetic field ($\mathcal{P}_{B_0}^{1/2}$) measured in unit of Gauss}\\
 \cline{2-5}
&\multicolumn{1}{c|}{$n_s$} &\multicolumn{3}{c|}{Scale invariant electric field $(n=2)$}\\
 \cline{3-5}
 &&\quad$10^{-18}$ G&\quad$10^{-20}$ G &\quad$10^{-22}$ G \\
 \hline
 \qquad $\omega_{eff}$&\qquad $0.9722$&\quad$0.32725$&\quad $0.3264$&\quad $0.3253$\\
 \qquad $T_{re}$ (GeV)&&\quad$0.023$&\quad $2.28$&\quad $228.6$ \\
\hline 
 \end{tabular}
 \label{minimaltab1}
 \end{table} 
 \begin{table}[t!]
\caption{
Constraining reheating and inflationary parameters through inflationary magnetogenesis}
Minimal plateau model ($\phi_*=0.01 M_p$)\\[.1cm]
\begin{tabular}{|p{3cm}|p{4cm}|p{4cm}|p{4cm}|  }
\hline
 \qquad Parameters&\multicolumn{3}{c|}{Current observed value of the magnetic field ($\mathcal{P}_{B_0}^{1/2}$) measured in unit of Gauss}\\
 \hline
 &\multicolumn{3}{c|}{Scale invariant electric field $(n=2)$}\\
 \cline{2-4}
 &\qquad$10^{-18}$ G&\qquad$10^{-20}$ G &\qquad$10^{-22}$ G\\
\hline
  \qquad$n_s^{min}$&\qquad$0.972$&\qquad $0.9707$&\qquad $0.9693$\\
  \qquad$n_s^{max}$&\qquad$0.9722$&\qquad $0.9722$&\qquad $0.9722$\\
  \qquad $\omega_{eff}^{min}$&\qquad$0.3144$&\qquad $0.239$&\qquad $0.1743$\\
    \qquad$\omega_{eff}^{max}$&\qquad$0.32725$&\qquad $0.3264$&\qquad $0.3253$\\
    \qquad$T_{re}^{min}$(GeV)&\qquad$10^{-2}$&\qquad $10^{-2}$&\qquad$10^{-2}$\\
    \qquad$T_{re}^{max}$(GeV)&\qquad$0.023$&\qquad $2.3$&\qquad$228.6$\\
    \hline
 \end{tabular}
 \label{minimaltab2}
 \end{table} 
In this section, we will introduce a special class of the inflationary model, the minimal plateau models proposed in \cite{Maity:2019ltu}. The potential of this type of model is a non-polynomial modification to the simple power-law potential $\phi^n$ and is given by 
 \bea
V_{min}=\Lambda \frac{m^{4-p}\phi^{p}}{1+\left(\frac{\phi}{\phi_*}\right)^p}~~,
\eea
where m and $\phi_*$ are two mass scales. The parameter $\lambda$ and the scale $m$ are fixed from WMAP normalization \cite{Komatsu:2010fb}. Only even values of the index $p$ are taken, as was the case for the chaotic inflation model. The new scale $\phi_*$ can be shown to fix the scalar spectral index and the scalar-to tensor ratio within the observational limit from Planck \cite{Ade:2015lrj,Akrami:2018odb}. For the numerical purpose, we choose $\phi_*=0.01 M_p$.\\ 
Before we go to the quantitative discussion to acquire one to one correspondence between the reheating parameters ($\omega_{eff} ~(\omega_\phi), T_{re}$) with $\mathcal{P}_{B_0}^{1/2}$, let us point out the usual inflationary parameters for this class of minimal models we discussed. The inflationary parameters $N_k$ and $r_k$ can be written as,
\bea
r_k=\frac{8M_p^2p^2}{\phi^2\left(1+\left(\frac{\phi}{\phi_*}\right)^p\right)^2}~~,~~N_k=\int\limits_{\phi_k}^{\phi_{end}}-\frac{\phi\left(\phi_*^p+\phi^p\right)}{p M_p^2 \phi_*^p }d\phi~~.
\eea  
Moreover, similar to the other inflation model, the initial conditions to solve the differential equation in the case of the perturbative reheating model are set as
\bea
\Phi(A=1)=\frac{3}{2}\frac{V_{end}}{m_{\phi}^4}~~,~~R(A=1)=0~,
\eea
where
\bea
V_{end}=\frac{m^{4-p}\phi_{end}^p}{1+\left(\frac{\phi_{end}}{\phi_*}\right)^p}~~,~~m=\left(\frac{3\pi^2M_p^4 r_k A_s}{2\Lambda \phi_k^p}\left(1+\left(\frac{\phi_k}{\phi_*}\right)^p\right)\right)^{\frac{1}{4-p}}~~.
\eea
We set $\Lambda=1$.\\
Details constraints on the reheating parameter space can be read from Figures (\ref{n2}), (\ref{perturbativedifferentw}) and tables (\ref{minimaltab1}),(\ref{minimaltab2}), (\ref{perturbativebound}).\\
It it interesting to observe that for minimal-inflation model, the reheating parameters in terms of scalar spectral index and large scale magnetic field posses both qualitatively and quantitatively different  compated to that of the $\alpha-$attractor model both quantitatively and quantitatively shown in the Fig.(\ref{perturbativedifferentw}). This difference can be attributed to the vary nature of the potential near its minima and the initial condition at the beginning of reheating. The numerical values and their constraints of reheating parameters ($\omega_{eff},T_{re}$) for different sample values of $\mathcal{P}_{B_0}^{1/2}$ cane be read from the table-\ref{minimaltab2} for instantaneous reheating case, and table-\ref{perturbativebound}, third panel of Fig.(\ref{perturbativedifferentw}) for perturbative reheating case.
\begin{table}[t!]
	\caption{Different inflationary models and their associated bounds on reheating temperature ($T_{re}$) and present magnetic field ($\mathcal{P}_{B0}^{\frac{1}{2}}$), measured in units of Gauss (Perturbative reheating dynamics)}
	 \begin{tabular}{|p{3cm}|p{2.5cm}|p{2.5cm}|p{2.5cm}|p{2.5cm}|}
\hline
~~Parameters&\multicolumn{2}{c|}{$\alpha$-attractor model ($\alpha=1$)}&\multicolumn{2}{c|}{Minimal plateau model $(\phi_*=0.01M_p$)}\\
\cline{2-5}
~&\qquad$\omega_\phi=0.5$&\qquad$\omega_\phi=2/3$&\qquad$\omega_\phi=0.5$&\qquad$\omega_\phi=2/3$\\
\hline
~~$\mathcal{P}_{B0}^{\frac{1}{2}}$ (minimum)&\qquad$ 10^{-22}$&\qquad$10^{-22}$&\qquad$ 10^{-22}$&\qquad$ 10^{-22}$\\
~~$\mathcal{P}_{B0}^{\frac{1}{2}}$ (maximum)&\qquad$3.4\times 10^{-14}$&\qquad$2.9\times 10^{-11}$&\qquad$6.0\times 10^{-16}$&\qquad$2.3\times 10^{-13}$\\
~~$n_s^{min}$&\qquad $0.9662$&\qquad $0.9669$&\qquad $0.9687$&\qquad $0.9678$\\
~~$n_s^{max}$&\qquad $0.9672$&\qquad $0.9688$&\qquad $0.9694$&\qquad $0.9694$\\
~~$T_{re}^{min}$ (GeV)&\qquad$10^{-2}$&\qquad$10^{-2}$&\qquad$10^{-2}$&\qquad$10^{-2}$\\
~~$T_{re}^{max}$ (GeV)&\qquad$2.4\times10^{4}$&\qquad$2.1\times10^{6}$&\qquad$1.7\times10^{3}$&\qquad$2.3\times10^{4}$\\
\hline
\end{tabular}
\label{perturbativebound}
\end{table}
\subsection{Constraining magnetogensis model: maximum possible value of $n$ and corresponding reheating temperature $T_{re}$}

\begin{table}[t!]
	\caption{Different inflationary models and their associated bounds on reheating temperature ($T_{re}$) and present magnetic field ($\mathcal{P}_{B0}^{\frac{1}{2}}$) (measured in units of Gauss ), considering maximum allowed value of the coupling parameter $n_{max}$}
  \begin{tabular}{|p{3cm}|p{2cm}|p{2cm}|p{2cm}|p{2cm}|p{2cm}|}
\hline
~Parameters &\multicolumn{2}{c|}{Axion $(f=50 M_p)$} &\multicolumn{2}{c|}{Higgs-Starobinsky}&\multicolumn{1}{c|}{Minimal plateau$(\phi_*=0.01M_p$)}\\
\hline
~&$n_s^{max}(0.9654)$&\quad   $n_s^{central}$&\multicolumn{1}{c|}{$n_s^{max}(0.9653)$}& \quad $n_s^{central}$ &\multicolumn{1}{c|}{$n_s^{max}(0.9722$)}\\
\hline

~$n_{max}$& ~$2.136$&~$2.138$&~$2.142$&~$2.144$ &\multicolumn{1}{c|}{$2.179$}\\
~$\mathcal{P}_{B0}^{\frac{1}{2}}$ (minimum)&~$10^{-22}$&~$10^{-22}$&~$10^{-22}$&~$10^{-22}$&\multicolumn{1}{c|}{$10^{-22}$}\\
~$\mathcal{P}_{B0}^{\frac{1}{2}}$ (maximum)&~$2.6\times 10^{-13}$&~$5.5\times 10^{-14}$&~$4.2\times 10^{-14}$&~$1.3\times 10^{-14}$&\multicolumn{1}{c|}{$2.3\times 10^{-14}$}\\
~$T_{re}^{min}$ (GeV)&~$ 10^{-2}$&~$10^{-2}$&~$10^{-2}$&~$10^{-2}$&\multicolumn{1}{c|}{$10^{-2}$}\\
~$T_{re}^{max}$ (GeV)&~$2.3\times10^{7}$&~$5.8\times10^{6}$&~$4.9\times10^{6}$&~$1.4\times10^{6}$&\multicolumn{1}{c|}{$2\times10^{6}$}\\
~$\omega_{eff}^{min}$ &~ $0.3281$&~ $0.2762$&~ $0.3297$&~$0.2904$&\multicolumn{1}{c|}{$0.3218$}\\
~$\omega_{eff}^{max}$&~$0.3309$&~$0.3042$&~$0.3315$&~$0.3120$&\multicolumn{1}{c|}{$0.3274$}\\
\hline
  \end{tabular}\\[.2cm]
  
  \label{maxntre}
\end{table}
Scaling properties of the large scale magnetic field as a function of momentum is not particularly well understood from the observation point view. Through out the paper, though, we have emphasized two special cases of interest related to  scale invariant electric field ($n=2$) and scale invariant magnetic field ($n=3$). Any other value of $n$ could have observational relevance in the near future. Keeping this motivation in mind, in this section we study the possible range of $n$ values which can give rise to observable strength of the magnetic field.  As has already been described before, the direct constraint on the value of $n$ will come from the strong coupling and the backreaction problem leading to  the fact that $n$ should lie within $0\leq n\leq n_{max}$, where $n_{max}$ can be determined from Eq. \ref{backr}, \ref{backr2}. Our primary goal of this section is to pin point the value of $n_{max}$ for different models under consideration. From the Fig.\ref{n}, we can clearly see that the permissible range of $n_{max}$ for various inflationary models should lie approximately within ($2.15 - 2.3$). 
\begin{figure}[t]
 	\begin{center}
 		\includegraphics[width=009.00cm,height=6.3cm]{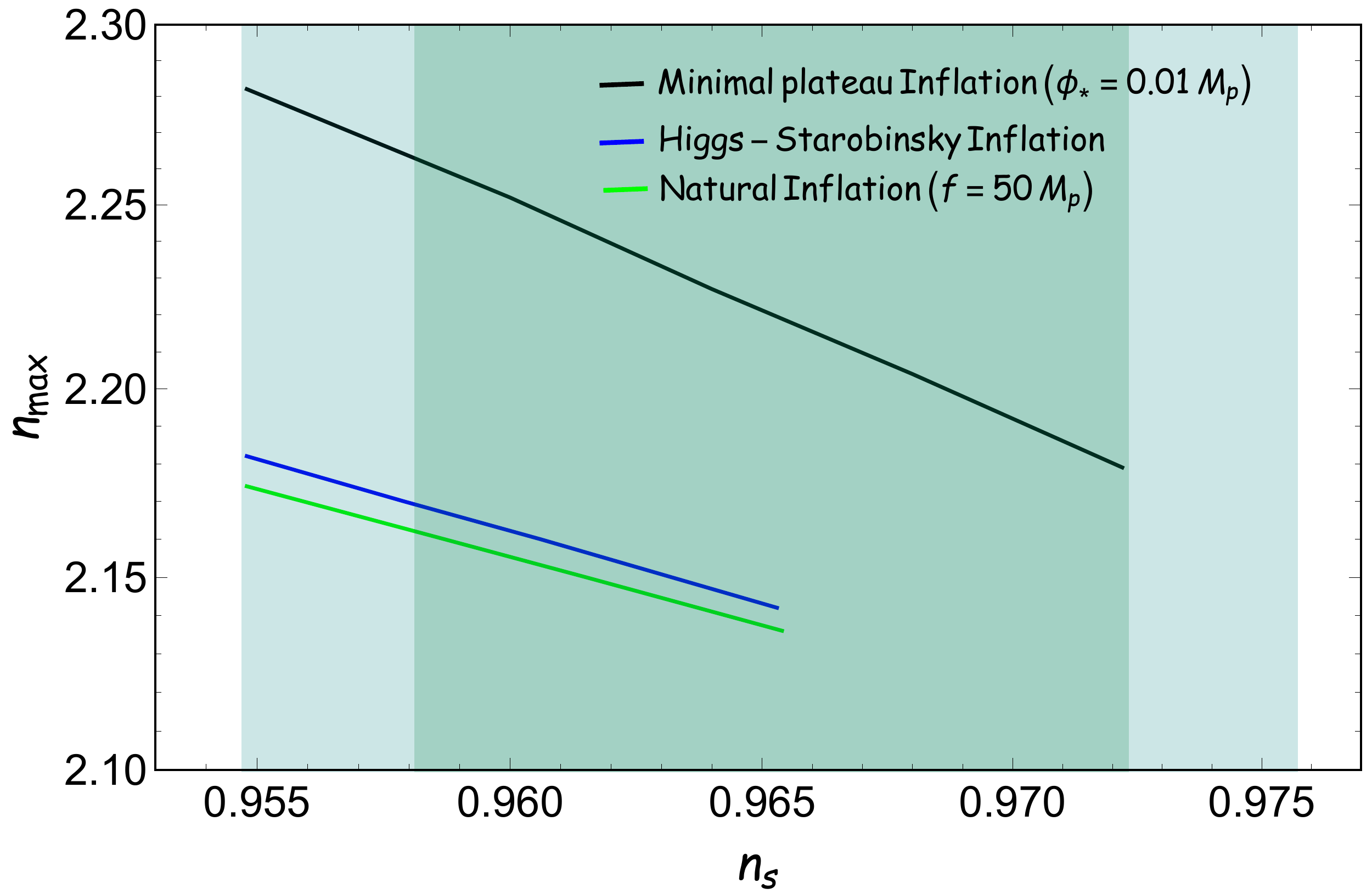}
 		 		\caption{We plot the variation of maximum value of $n$ ($n_{max}$) as a function of spectral index within 2$\sigma$ range, $n_s$ from Planck \cite{Akrami:2018odb} for different inflationary model. In this plot the maximum values of the spectral index correspond to the instantaneous reheating condition $N_{re}\to0$. }
 		\label{n}
 	\end{center}
 \end{figure}
As has been observed before, conventional magnetogenesis scenarios are generically viable for low scale inflationary models or in other words low reheating temperature regime. This turned to be untrue for higher $n\sim n_{max}$. In order to calculate reheating temperature associated with the maximum possible value of the $n$, we consider two sample value of the spectral index $n_s=(n_s^{central}=0.9649, n_s^{max})$.  Irrespective of the inflationary models under consideration, reheating temperature can be observed as high as $\sim$ $10^{7}$ GeV without any backreaction problem. In table \ref{maxntre}, we have provided the limiting values of the $(\omega_{eff}, T_{re}, \mathcal{P}_{B_0}^{1/2})$ for two different values of the scalar spectral index $n_s$. To this end, let us point out an important result of our analysis that the minimal plateau model can be observationally discarded as it predicts reheating temperature less than BBN limit ( $10^{-2}$ Gev ) for $0\leq \omega_{eff}\leq \frac{1}{3}$.
 
\section{Summary and discussion:}

Among all magnetogenesis scenarios, the inflationary magnetogenesis is well-motivated and simple mechanism to explain the origin of the large scale magnetic field in our universe. Inflation helps boosting small scale magnetic field born out of quantum vacuum into cosmological scale through Faraday's law of electromagnetic induction. In the present paper we show how the same mechanism can play profound role during reheating phase of the early universe as well.
Primary assumption behind this mechanism
to act during reheating is the negligible electrical conductivity.
As a result strong electric field can survive during this phase and is dynamically converted into magnetic field.
Hence the resulting primordial magnetic field energy density does not evolve as radiation energy density ($1/a^4$) during this phase, instead evolve as $\propto 1/H^2 a^{6}$ till the end of the reheating phase. After the end of the reheating electric field dies out fast due to the large electric conductivity and comoving magnetic energy density freezes out until today. Considering this physical effect, the present-day amplitude of the magnetic field originated during inflation becomes observationally viable with specific set of reheating parameters. Our motivation of this work is to figure out these constraints on the reheating parameters which in turn can constraint inflation and magnetogeneis models.\\
We have derived explicit connection among the infltionary $(n_s)$, reheating ($\omega_{eff}, T_{re}$), and magnetogenesis $(n)$ model parameters assuming two different reheating scenarios. Large-scale magnetic field gives rise to a stringent constraints on both the reheating parameters depending upon the models under considerations. Considering both backreaction and strong coupling problems in to account, the maximum allowed value of $n$ would be  $n_{max} \sim 2.3$ for all the models such as natural inflation and Higgs-Starobinsky, and minimal plateau model. The associated maximum allowed value of $n$, the reheating temperature has been shown to lie within a wide range of ($10^{-2},10^7$) GeV.

We also have some definite prediction of our analysis once we fix the model and the large scale magnetic field. For reheating scenario with constant equation of state $\omega_{eff}$, once we fixed the scalar spectral index, a unique value of $\omega_{eff}$ is predicted associated with the specific choice of the present-day magnetic field. This unique $\omega_{eff}$, further predicts a specific value of reheating temperature $T_{re}$. The quantitative values of those parameters for $n=2$ magnetogenesis model are provided in the tables \ref{axiontab1},\ref{alphatab1} and \ref{minimaltab1}.  As reheating and inflationary parameters are connected, the present-day strength of the magnetic field also provides possible limits on the inflationary scalar spectral index.  This extra bound from inflationary magnetogenesis essentially narrows down the possible value of $n_s$ within the $1\sigma$ range of $n_s=0.9649\pm0.0042$ ($68 \%$ CL, Planck TT,TE,EE+lowE+lensing) from Planck \cite{Akrami:2018odb}. (shown in tables \ref{axiontab2}, \ref{alphatab2}, and \ref{minimaltab2} for different inflationary models). Model independent constraint on the effective equation of state is obtained as $(0.15< \omega_{eff}<\frac{1}{3})$. One of the main drawbacks of describing the reheating dynamics by a constant $\omega_{eff}$ is that it does not enable us to decode the nature of the inflation potential near its minimum. This drawback motivated us to consider the reheating in the perterbative framework where inflation equation of state is described by $\omega_{\phi} \equiv (p-2)/(p+2)$ with inflaton potential $V(\phi) \sim \phi^{p}$ near its minimum.   

 For the perturbative reheating scenario the range of inflaton equation of state $ 0 <\omega_\phi \lesssim 0.28$ is found to be observationally not viable for ~$\mathcal{P}_{B0}^{\frac{1}{2}} \gtrsim 10^{-18}$ G. Increasing the lower limit of the magnetic field will futher widen the non-viable range of  $\omega_{\phi}$. We think it is extremely interesting in the context of constructing inflaton model building. This observation provides us a strong constraint on the possible form of the inflaton potential near its minium with $p \gtrsim 3.6$ considering the aforesaid observable limit of the present day magnetic field strength. Furhtermore, we if one consideres the highrer equation of state the high scale inflation model with reheating temeperature as high as $10^7$ GeV is possible. The main outcomes of the perturbative analysis for different inflationary model shown in table-\ref{perturbound2}, \ref{perturbativebound}.}
 
 As we already mentioned, our primary motivation is to give a proper methodology to probe the reheating phase which follows the standard inflationary era through the present-day large-scale magnetic field combined and the CMB anisotropy. In order to do that, we consider the simplest magnetogenesis scenario the so called Ratra's model. Different magnetogenesis model can be studied straightforwardly following the formalsim devolpoed in this paper to probe reheating phase, as has already been applied recently in papers \cite{Maity:2021qps,Bamba:2020qdj}. Only differences may arise into the allowed reheating and inflationary parameters space. However, to distinguish those models, we need more physical observable such as primordial gravitational wave \cite{Sharma:2021rot}, dark matter, etc. Our eventual plan in the future is to obtain some universal constraints on the reheating and inflationary parameters taking into account aforesaid observables.
 \section{ACKNOWLEDGEMENT}
 We would like to thank the HEP and Gravity groups at IIT Guwahati for useful discussions. We thank the anonymous referees for
useful comments which helps us to improve our work.
 \hspace{0.5cm}
  
\end{document}